\documentclass[fleqn,usenatbib]{mnras}

\usepackage{newtxtext,newtxmath}

\usepackage[T1]{fontenc}

\DeclareRobustCommand{\VAN}[3]{#2}
\let\VANthebibliography\thebibliography
\def\thebibliography{\DeclareRobustCommand{\VAN}[3]{##3}\VANthebibliography}

\usepackage{graphicx}	
\usepackage{amsmath}	
\usepackage{fontawesome}    
\usepackage{anyfontsize}    

\title[\textsc{Psi-GAN}]{\textsc{Psi-GAN}: A power-spectrum-informed generative adversarial network for the emulation of large-scale structure maps across cosmologies and redshifts}

\author[P. Bhambra et al.]{
Prabh Bhambra,$^{1}$\thanks{E-mail: prabh.bhambra.12@ucl.ac.uk}
Benjamin Joachimi,$^{1}$
Ofer Lahav,$^{1}$
Davide Piras$^{2,3}$
\\
$^{1}$Department of Physics and Astronomy, University College London, Gower Street, London WC1E 6BT, UK
\\
$^{2}$Centre Universitaire d’Informatique, Université de Genève, 7 route de Drize, 1227 Genève, Switzerland
\\
$^{3}$Département de Physique Théorique, Université de Genève, 24 quai Ernest Ansermet, 1211 Genève 4, Switzerland
}

\date{Accepted XXX. Received YYY; in original form ZZZ}

\pubyear{2024}

\begin{document}
\label{firstpage}
\pagerange{\pageref{firstpage}--\pageref{lastpage}}
\maketitle

\begin{abstract}
Simulations of the dark matter distribution throughout the Universe are essential in order to analyse data from cosmological surveys. $N$-body simulations are computationally expensive, and many cheaper alternatives (such as lognormal random fields) fail to reproduce accurate statistics of the smaller, non-linear scales. In this work, we present \textsc{Psi-GAN} (\textbf{P}ower-\textbf{s}pectrum-\textbf{i}nformed \textbf{G}enerative \textbf{A}dversarial \textbf{N}etwork), a machine learning model which takes a two-dimensional lognormal dark matter density field and transforms it into a more realistic field. We construct \textsc{Psi-GAN} so that it is continuously conditional, and can therefore generate realistic realisations of the dark matter density field across a range of cosmologies and redshifts in $z \in [0, 3]$. We train \textsc{Psi-GAN} as a generative adversarial network on $2\,000$ simulation boxes from the Quijote simulation suite. We use a novel critic architecture that utilises the power spectrum as the basis for discrimination between real and generated samples. \textsc{Psi-GAN} shows agreement with $N$-body simulations over a range of redshifts and cosmologies, consistently outperforming the lognormal approximation on all tests of non-linear structure, such as being able to reproduce both the power spectrum up to wavenumbers of $1~h~\mathrm{Mpc}^{-1}$, and the bispectra of target $N$-body simulations to within ${\sim}5$ per cent. Our improved ability to model non-linear structure should allow more robust constraints on cosmological parameters when used in techniques such as simulation-based inference.
\end{abstract}




\section{Introduction}
\label{sec:intro}

The standard model of cosmology, known as $\Lambda$CDM \cite[see e.g.][]{peebles}, describes a Universe consisting of cold dark matter (CDM), ordinary matter (baryons), and includes the existence of a cosmological constant $\Lambda$ associated with dark energy. $\Lambda$CDM favours that the relative abundance of dark matter is approximately five times that of baryonic matter, making it the predominant form of matter throughout the Universe \citep{cdm_review}. The model describes a Universe in which galaxies form along and trace the cosmic web structure formed by dark matter, consisting of filaments which connect clusters and surround voids. Although the gravitational effects of dark matter have been observed in many different ways, the nature of dark matter itself remains a mystery \cite[see e.g.][and references therein]{cdm_search}.

$N$-body simulations are a common tool used to analyse the origin and evolution of the cosmic web structure formed by dark matter \cite[see e.g.][]{n_body_techniques, galaxy_quasar_sim, gadget_2, millennium_ii, quijote, camels, gadget_4}. In its simplest form, running an $N$-body simulation involves initialising a number of massive particles in a cubic box of fixed comoving dimensions, imposing periodic boundary conditions, and then allowing gravity to act on the particles through its gravitational potential \cite[governed by the Poisson equation;][]{gadget_4}. The initial conditions of the $N$-body simulation are often approximated by a Gaussian random density field and, starting from these initial conditions, the positions and velocities of each particle are updated iteratively over a series of timesteps until today ($z=0$).

There exist many different implementations of $N$-body simulations with differing complexity and accuracy. Direct methods, in which the force on each particle with respect to every other particle is calculated for each timestep, are extremely computationally expensive, and so approximations are used to reduce the time taken to run a simulation. These approximations include: tree code methods \citep{tree_code}, fast-multipole methods \citep{fast_multipole}, particle-mesh methods \citep{particle_mesh}, adaptive mesh refinement \citep{adaptive_mesh_refinement, enzo_code}, and combinations such as \textsc{Tree-PM} \cite[see e.g.][]{gadget_2, gadget_4}. Despite the improvements in speed due to these approximations, $N$-body simulations are still computationally expensive to run and require access to high-performance computing systems. The time and computing resources required to run a sufficient number of $N$-body simulations limits our ability to study the nature of dark matter and the Universe through techniques such as simulation-based inference \citep{sbi}.

When a significantly large number of simulations is required, it is common to resort to cheaper approximations. One such approximations for describing dark matter fields is to use a lognormal random field \cite[see e.g.][]{lognormal_model, lognormal_fourier, flask, lognormal_des, glass}. A lognormal random field can be easily obtained from a given Gaussian random field, and can be entirely described by very few parameters: the mean $\mu$ and variance $\sigma^{2}$ of the associated Gaussian random field, and a shift parameter $\lambda$. A lognormal random field also demonstrates a skew, which is useful in modelling the matter overdensity field given that it varies from values of $-1$ in voids to values in the range of ${\sim}10^{7}$ in clusters. These properties make lognormal random fields a useful approximation of the matter overdensity field. However, as discussed in \citet{flask} and \citet{glass}, its low computational complexity comes with limitations. Lognormal random fields are able to reproduce a power spectrum to a high level of accuracy as the power spectrum relies only on the amplitudes of Fourier modes. However, they are unable to reproduce accurate statistics that rely on the phases of Fourier modes, which contain much of the information regarding non-linear structure \citep{phase_information}.

Recently, machine learning (ML) methods have been used to approximate $N$-body simulations. \citet{cosmic_web_gan} and \citet{cosmogan} used generative adversarial networks \cite[GANs;][]{gan} to emulate slices of $N$-body simulations and weak lensing convergence maps, respectively. \citet{cosmo_3d_1} and \citet{cosmo_3d_2} extended this approach from two-dimensional slices to three-dimensional simulation boxes, and showed that GANs are able to reproduce the large-scale and small-scale features of $N$-body simulations. \citet{cosmo_unet_1} and \citet{cosmo_unet_2} trained U-Nets \citep{unet} to learn the non-linear growth of cosmic structure.

More recently, \citet{fast_nbody_sims} used a U-Net in a GAN framework to emulate $N$-body simulations by learning how to transform a corresponding lognormal approximation. \citet{lensing_cyclegan} similarly used a U-Net in a Cycle GAN \cite[an unpaired image-to-image method;][]{cyclegan} framework to learn unpaired translation from lognormal approximations of weak lensing mass maps to non-Gaussian counterparts. \citet{gansky} developed new network layers in order to generate full-sky weak lensing mass maps from lognormal approximations.

While useful, very few methods consider the impact of cosmology and redshift on the structure of the cosmic web. \citet{fast_nbody_sims} considered cosmology and redshift dependence for a simplified low-resolution case, however this dependence was not built into the model. \citet{cosmology_dependence} encode cosmology dependence into their U-Net-based model to output non-linear displacements and velocities of $N$-body simulation particles based on their linear inputs.

In this paper we aim to improve lognormal approximations through the use of ML techniques, across a range of cosmologies and redshifts. We build upon the work of \citet{fast_nbody_sims} by extending their approach to fully capture cosmology and redshift dependence, with the long-term goal of integrating our work into \textsc{Glass} \citep{glass}. Our approach starts from the Quijote $N$-body simulation suite \citep{quijote}, which contains $2\,000$ $N$-body simulation boxes with cosmologies sampled from a five-dimensional Latin hypercube. The simulation suite includes snapshots at five redshifts as well as the initial conditions, which we use to create a dataset of pairs of lognormal and $N$-body slices. We train a conditional U-Net in a GAN in order to learn an image-to-image translation between the domains. Our novel method uses the power spectrum of the generated emulation to inform the network during training and guide it towards reproducing the structure of $N$-body simulations across all scales.

Our paper is structured as follows. In Section~\ref{sec:data} we describe the data used from the Quijote simulations. In Section~\ref{sec:method} we describe the data generation procedure used to obtain a corresponding lognormal slice for each $N$-body simulation slice, our model architecture, as well as our training, validation, and testing methods. In Section~\ref{sec:results} we present the results of our method, including evaluating model performance within the domain of the training data, as well as testing its ability to interpolate within the cosmology and redshift spaces. We conclude in Section~\ref{sec:conc} with a summary of our work, as well as suggestions for future work needed to meet our long-term goal of \textsc{Glass} integration.

\section{Data: simulations and matter fields}
\label{sec:data}

In this work, we use the Quijote simulation suite \citep{quijote}. We specifically use simulations from the Latin hypercube, in which the values of the matter density parameter ($\Omega_{\mathrm{m}}$), the baryon density parameter ($\Omega_{\mathrm{b}}$), the Hubble parameter ($h$), the scalar spectral index ($n_{\mathrm{s}}$), and the root mean square of the matter fluctuations in spheres of radius $8~h^{-1}~\mathrm{Mpc}$ ($\sigma_{8}$) are varied by sampling from a five-dimensional Latin hypercube. We only consider massless neutrinos, and a constant value for the dark energy equation of state parameter $w = -1$ (i.e. a constant $\Lambda$). This Latin hypercube contains $2\,000$ standard simulations, each containing $512^{3}$ dark matter particles in a box with comoving length of $1\,000~h^{-1}~\mathrm{Mpc}$. The limits of the Latin hypercube are shown in Table~\ref{tab:latin_hypercube} along with the corresponding fiducial values. We utilise both the initial conditions at $z = 127$ of each simulation, as well as snapshots at redshifts $z \in \{0, 0.5, 1, 2, 3\}$, thus forming a dataset spanning a range of cosmologies and redshifts.

\begin{table}
    \centering
    \caption{The limits and fiducial values for each cosmological parameter in the Quijote simulation's Latin hypercube suite.}
    \label{tab:latin_hypercube}
    \begin{tabular}{ccc}
        \hline
        Parameter               & Limits                        & Fiducial Value    \\
        \hline
        $\Omega_{\mathrm{m}}$   & $\left[  0.1,0.5  \right]$    & $0.3175$          \\
        $\Omega_{\mathrm{b}}$   & $\left[ 0.03,0.07 \right]$    & $0.049$           \\
        $h$                     & $\left[  0.5,0.9  \right]$    & $0.6711$          \\
        $n_{\mathrm{s}}$        & $\left[  0.8,1.2  \right]$    & $0.9624$          \\
        $\sigma_{8}$            & $\left[  0.6,1.0  \right]$    & $0.834$           \\
        \hline
    \end{tabular}
\end{table}

For each simulation, we convert the particles' positional information to a continuous field through a mass assignment scheme. Throughout this work, we will consider the matter overdensity field $\delta(\mathbf{x})$, defined as:
\begin{equation}
    \delta(\mathbf{x}) = \frac{\rho(\mathbf{x})}{\Bar{\rho}} - 1,
    \label{eq:overdensity}
\end{equation}
where $\rho(\mathbf{x})$ is the matter density at each position $\mathbf{x}$, and $\Bar{\rho}$ is the mean density in the simulation box.

We consider a three-dimensional regular grid with $N^{3} = 512^{3}$ voxels. The interpolation of the overdensity field over the grid is then obtained by evaluating the continuous function,
\begin{equation}
    \Tilde{\delta}(\mathbf{x}) = \int \frac{\mathrm{d}^{3}\mathbf{x'}}{(2 \pi)^{3}} W(\mathbf{x} - \mathbf{x'}) \delta(\mathbf{x'}),
    \label{eq:interpolation}
\end{equation}
where $W(\mathbf{x})$ is the weight function which describes the number of grid points, per dimension, to which each particle is assigned. We utilise the piecewise cubic spline interpolation scheme \citep{highorder_interpolation, pcs_interpolation} in which the weight function is symmetric, positively defined, and separable such that $W(\mathbf{x}) = W_{\mathrm{1D}}(x_{1}/H) W_{\mathrm{1D}}(x_{2}/H) W_{\mathrm{1D}}(x_{3}/H)$, with $H$ being the grid spacing, and $W_{\mathrm{1D}}$ being the unidirectional weight function:
\begin{equation}
    W_{\mathrm{1D}}(s) =
    \begin{cases}
        \frac{1}{6}(4 - 6s^{2} + 3 \lvert s \rvert ^{3})    & \textrm{if}~0 \leq \lvert s \rvert < 1, \\
        \frac{1}{6}(2 - \lvert s \rvert )^{3}               & \textrm{if}~1 \leq \lvert s \rvert < 2, \\
        0                                                   & \textrm{otherwise}.
    \end{cases}
    \label{eq:pcs}
\end{equation}

\section{Method}
\label{sec:method}

The goal of this work is to be able to train a model that can transform two-dimensional lognormal overdensity fields into more realistic overdensity fields with statistics that match those of the Quijote Latin hypercube across redshifts and cosmologies. In order to do this, we first create a dataset containing pairs of two-dimensional slices of the Quijote Latin hypercube and their corresponding lognormal counterpart (Section~\ref{subsec:method_data}), we then train a machine learning model to apply this transformation (Sections~\ref{subsec:method_model}~and~\ref{subsec:method_training}), and finally validate the model using a set of statistical metrics (Section~\ref{subsec:method_val_test}).

\subsection{Data generation}
\label{subsec:method_data}

In order to create the required dataset, we obtain $n = 16$ slices for each three-dimensional simulation box by slicing each box along a chosen axis such that each slice has a depth of $32$ pixels. We reduce the dimensions of the slices from three to two by taking the depth-wise mean. The depth of each slice is then given by $1\,000/n~h^{-1}~\mathrm{Mpc} = 62.5~h^{-1}~\mathrm{Mpc}$, which was chosen to be lower than the approximate depth of matter shells in \textsc{Glass} \citep{glass_shell_thickness}. A shallower depth ensures that more small-scale structure remains in the slices, thus making it more difficult to model. Successfully reproducing slices of this depth, will ensure that \textsc{Psi-GAN} will also be able to reproduce slices of a greater depth. While the matter shells in \textsc{Glass} have varying depth, we leave incorporating this depth dependence into the model to future work. Our training data spans all of the $2\,000$ cosmologies in the Quijote Latin hypercube at redshifts of $z \in \{0, 0.5, 1, 2, 3\}$, resulting in $16 \times 2\,000 \times 5 = 160\,000$ slices.

In order to generate corresponding lognormal counterpart to each slice we follow the procedure outlined by \citet{fast_nbody_sims}. While a brief description will be provided here we direct the reader to \citet{fast_nbody_sims} for a more detailed description of this procedure.

We start by measuring the two-dimensional power spectrum of each slice $P(k)$. In order to generate a lognormal random field with the given measured power spectrum, we follow \citet{lognormal_model} and \citet{lognormal_fourier}. We then convert $P(k)$ to the matter correlation function $\xi_{\mathrm{LN}}(r)$, and calculate the corresponding Gaussian correlation function:
\begin{equation}
    \xi_{\mathrm{G}} = \mathrm{ln} \left[ 1 + \xi_{\mathrm{LN}}(r) \right].
    \label{eq:correlation_function}
\end{equation}
We convert this Gaussian correlation function back to Fourier space to obtain a Gaussian power spectrum $P_{\mathrm{G}}(k)$.

A zero-mean Gaussian field is entirely defined by its power spectrum which depends only on the absolute values of the Fourier coefficients, therefore the Fourier phases can be uniformly sampled in the interval $[0, 2\pi)$ in order to create a realisation of a Gaussian random field \citep{phase_entropy_1, phase_entropy_2, phase_difference}. However, as we aim to generate Gaussian random fields $\delta_{\mathrm{G}}$ with high correlations to each given $N$-body slice, we instead use the set of phases from the corresponding slice of the initial conditions at $z=127$. The lognormal field $\delta_{\mathrm{LN}}$ is then calculated by evaluating
\begin{equation}
    \delta_{\mathrm{LN}} = \mathrm{exp}\left( \delta_{\mathrm{G}} - \sigma_{\mathrm{G}}^{2}/2 \right)
    \label{eq:lognormal_field}
\end{equation}
for each grid point, where $\sigma_{G}$ is the standard deviation of the Gaussian field. For these operations, we used the \textsc{Python} package \textsc{nbodykit} \citep{nbodykit}.

There are two limitations to this method due to the fact that we are measuring the power spectrum from a grid. Firstly, due to relying only on the simulation boxes for the measured power spectrum, we are only able to survey a limited range of $k \in [0.025, 1]~h~\mathrm{Mpc}^{-1}$. In order to access larger scales, we use \textsc{Class} \citep{class} to generate a theoretical power spectrum for $k \in [10^{-5}, 0.025]~h~\mathrm{Mpc}^{-1}$ and concatenate this with the measured power spectrum.

Secondly, we observe a discrepancy in the power spectrum of the generated lognormal field and the measured power spectrum from the Quijote slice. This can be attributed to correlations in phases being introduced when converting the Quijote initial conditions (obtained by second-order Lagrangian perturbation theory) to a density field. We correct for this discrepancy by iteratively re-scaling $P_{\mathrm{G}}(k)$, which is used to generate the lognormal field $\delta_{\mathrm{LN}}$. Each iteration involves generating a lognormal field from $P_{\mathrm{G}}(k)$ as per Equation~\ref{eq:lognormal_field}, measuring its power spectrum $P_{\mathrm{LN}}(k)$, calculating the ratio of $P_{\mathrm{LN}}(k)$ to the target power spectrum $P(k)$ at each $k$, and then rescaling $P_{\mathrm{G}}(k)$ by this ratio at each value of $k$. This process is iterated through until $P_{\mathrm{LN}}(k)$ matches the target power spectrum $P(k)$ to within a $0.1$ per cent discrepancy at all values of $k$.

We are left with a dataset of pairs of lognormal and Quijote slices ($\delta_{\mathrm{LN}}$ and $\delta_{\mathrm{NB}}$), which we split into a number of sets. We firstly reserve all slices across all redshifts of cosmologies $\#1586$ and $\#815$ (randomly selected) as part of the test set in order to test model performance on unseen cosmologies. As we only have snapshots at certain redshifts, we create an additional set of lognormal slices at redshift $z \in \{0.25, 0.75\}$ for cosmology $\#663$ (which we will refer to as our ``fiducial'' cosmology from now on, as it is the closest cosmology in our dataset to the Quijote fiducial cosmology) in order to test the model's ability to interpolate between redshifts. We follow the previously outlined procedure for producing these lognormal slices. However as we have no Quijote snapshots at these redshifts (and are therefore unable to measure a power spectrum), we create a “measured” power spectrum by linearly interpolating the power spectrum at each value of $k$ between redshift snapshots. Furthermore, we reserve 512 randomly chosen slices at each redshift as part of a test set to assess model performance on cosmologies and redshifts within the training set. 10 per cent of the remaining dataset is used for validation, with the other 90 per cent being used for training. Table~\ref{tab:datasets} summaries these six sets of data used in the training, validating, and testing of our model.

\begin{table*}
    \centering
    \caption{A summary of the datasets used in training (bottom division), validating (middle division), and testing (top division) our model.}
    \label{tab:datasets}
    \begin{tabular}{p{0.225\textwidth}p{0.3\textwidth}p{0.225\textwidth}p{0.15\textwidth}}
        \hline
        Set name & Description & Cosmology & Redshift \\
        \hline
        Interpolate cosmology test set $\#1$ & Testing on reserved cosmology, unseen during training process & Simulation $\#1586$ & $z \in \{0, 0.5, 1, 2, 3\}$ \\
        Interpolate cosmology test set $\#2$ & Testing on reserved cosmology, unseen during training process & Simulation $\#815$ & $z \in \{0, 0.5, 1, 2, 3\}$ \\
        Interpolate redshift test set & Testing interpolation between redshifts & Simulation $\#663$ & $z \in \{0.25, 0.75\}$ \\
        Randomly split test set & Testing within the domain of the training data using randomly chosen slices at each $z$ & Randomly selected from all simulations (excluding $\#1586$ and $\#815$) & $z \in \{0, 0.5, 1, 2, 3\}$ \\
        \hline
        Validation set & Validating the model using 10 per cent of the remaining slices not used for testing & All simulations (excluding $\#1586$ and $\#815$) & $z \in \{0, 0.5, 1, 2, 3\}$ \\
        \hline
        Training set & Training the mode with 90 per cent of the remaining slices not used for testing & All simulations (excluding $\#1586$ and $\#815$) & $z \in \{0, 0.5, 1, 2, 3\}$ \\
        \hline
    \end{tabular}
\end{table*}

\subsection{Model architecture}
\label{subsec:method_model}

\begin{figure*}
    \centering
    \includegraphics[width=\textwidth]{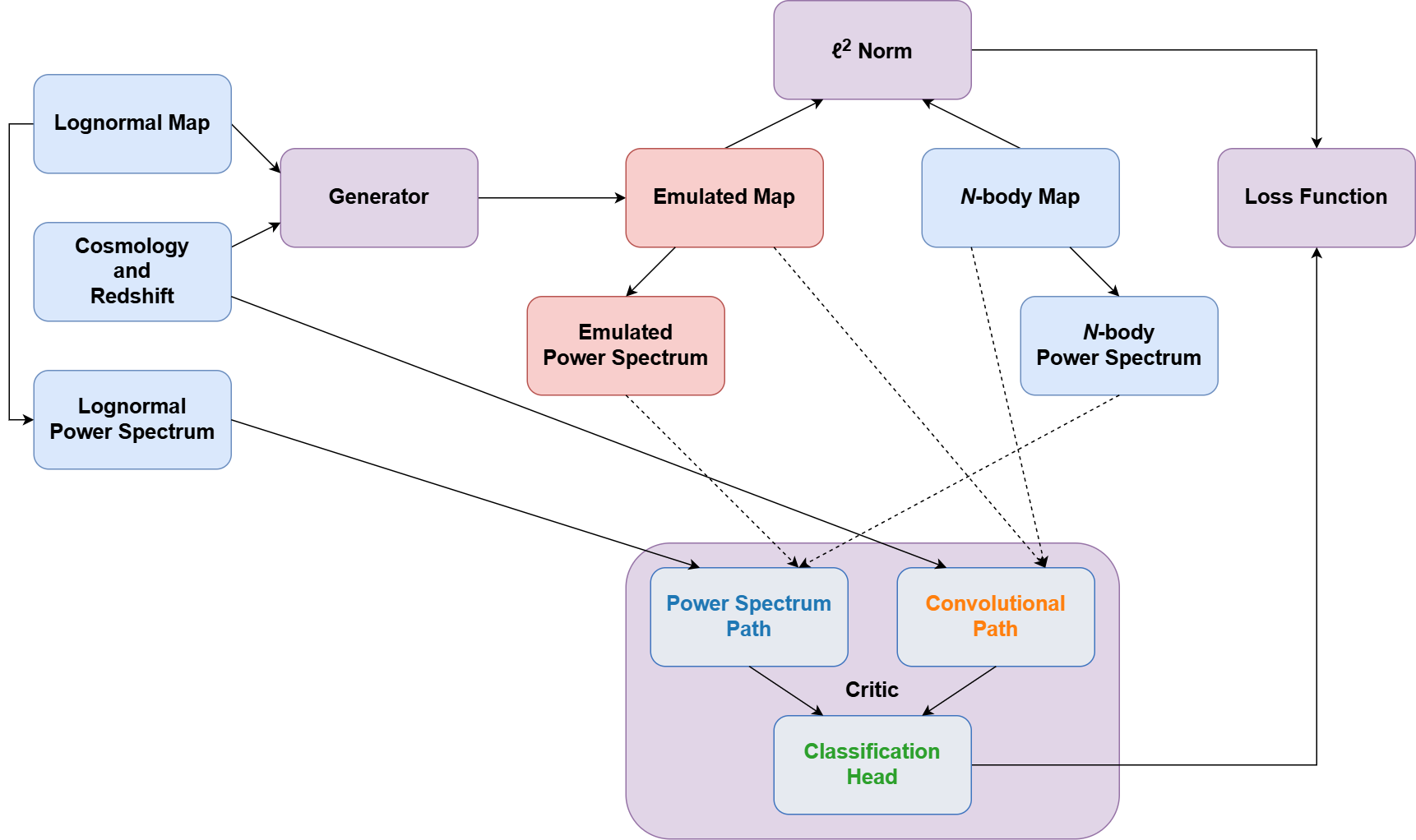}
    \caption{A schematic showing the full construction of the \textsc{Psi-GAN} framework. Data which are part of the initial dataset are coloured in blue, while data that are calculated by the \textsc{Psi-GAN} framework are coloured in red. All computational steps are coloured in purple. Note that the critic takes in either an emulated map or an $N$-body map as an input (along with their associated power spectrum), which is indicated by dashed lines. The internal structure of the critic is also shown, and can be seen in more detail in Figure~\ref{fig:critic}.}
    \label{fig:gan}
\end{figure*}

We train a Wasserstein GAN with gradient penalty \cite[WGAN-GP;][]{wgan, wgan_gp} consisting of a generator (${\sim}2.06 \times 10^{7}$ parameters) and a critic (${\sim}2.59 \times 10^{7}$ parameters). In a traditional GAN, the generator and critic are adversarially trained in tandem in order to produce generated data that is identical to real data. Our approach builds physics into the critic of the GAN to constrain the generator to produce data that is physically consistent with the target domain. A full schematic of the \textsc{Psi-GAN} framework can be found in Figure~\ref{fig:gan}. In this figure, we demonstrate how an emulation can be generated by feeding a lognormal density field, along with its associated cosmology and redshift, into the generator. This framework is trained via a loss function which depends on the output of our physics-informed critic, which takes as inputs either an emulated or $N$-body map, its associated power spectrum, its associated cosmology and redshift, and finally the power spectrum of the corresponding lognormal density field.

\begin{figure*}
    \centering
    \includegraphics[width=0.8\textwidth]{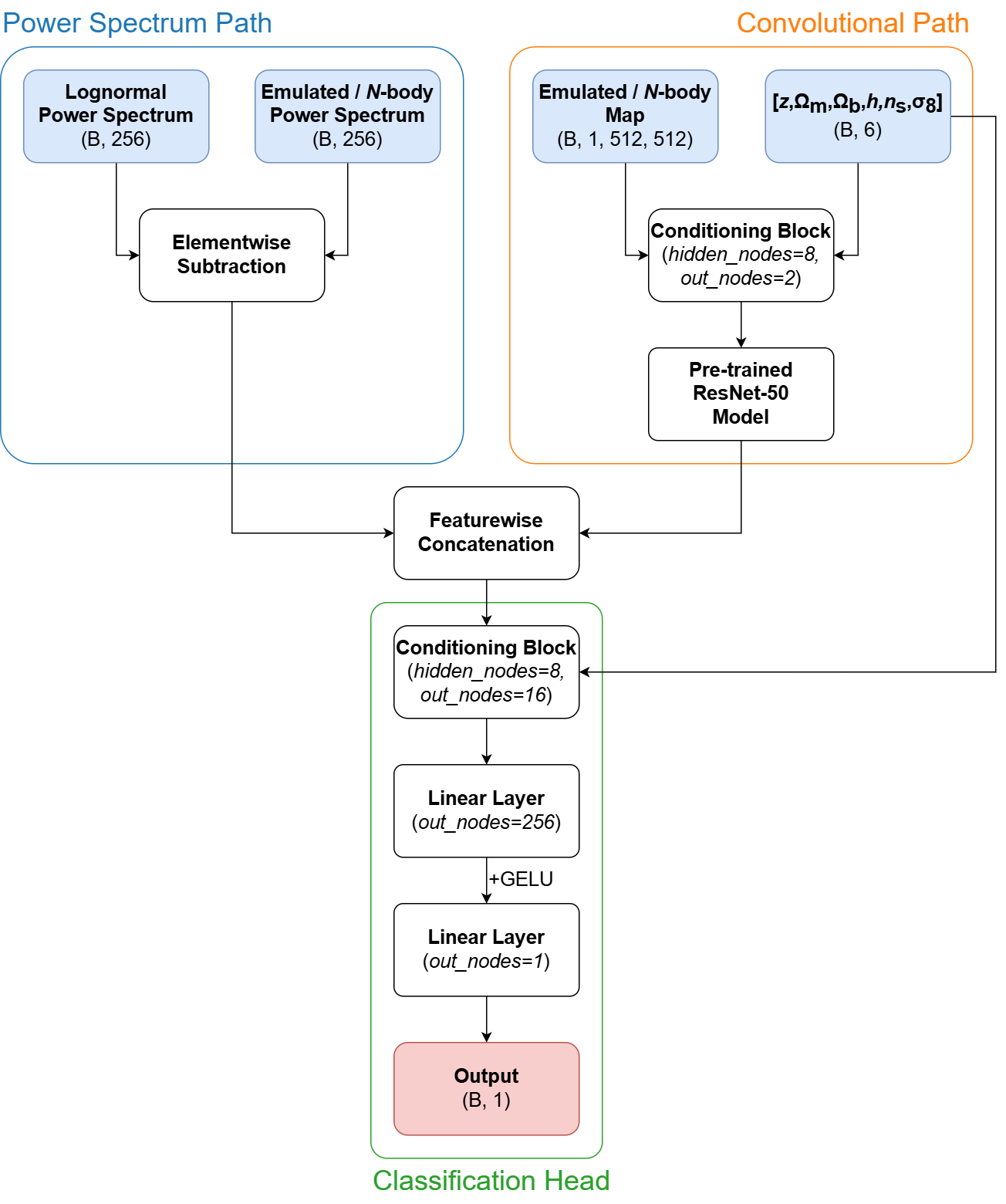}
    \caption{A schematic showing the construction of our critic. The power spectrum path, convolutional path, and classification head are outlined and labelled in blue, orange, and green, respectively. We also provide information, in parentheses, regarding the dimensions for both the inputs (blue) and outputs (red) in roman font, and the hyperparameters of each layer in italics. Dimensions are quoted in the ``batch, channels, $\ast$'' convention, where $\ast$ represents any number of latent dimensions and B is used as a placeholder for the batch dimension of all inputs and outputs. All convolutional layers use circular padding in order to maintain the height and width of the input, and GELU represents the Gaussian Error Linear Unit \citep{gelu}. The conditioning block is used to inform the network of the cosmology and redshift of the emulation, and is defined in Appendix~\ref{app:model}.}
    \label{fig:critic}
\end{figure*}

Details regarding the computation blocks used to construct \textsc{Psi-GAN} along with the construction of the generator itself can be found in Appendix~\ref{app:model}, while the construction of the critic is shown in Figure~\ref{fig:critic}. The critic consists of two paths, a convolutional path and a power spectrum path (shown in orange and blue, respectively). The convolutional path takes an input image (with cosmology and redshift embeddings) and processes it using a pre-trained ResNet-50 model \citep{resnet} to obtain a feature representation of the input.\footnote{A pre-trained ConvNeXt-T \citep{convnext} model was also investigated as an option, however this resulted in an increase in training time by a factor of ${\sim}2$. The initial results also indicated poor performance due to the ConvNeXt-T model immediately down-sampling the input by a factor of $4$, thus placing a limit on how well small-scale features can be backpropagated to the generator.} The power spectrum path takes the power spectrum of the input image and compares this to the power spectrum of the corresponding lognormal map via an elementwise subtraction. Both the feature representation and the power spectrum comparison are concatenated and then fed into a linear classifier, along with another set of cosmology and redshift embeddings.

GANs for image synthesis often use a purely feature-based network for the critic, however using a fully-convolutional critic resulted in the generator altering the power spectrum of its input when attempting to generate a more realistic output. The power spectrum path was then added to the critic in order to guide the generator towards not altering the power spectrum. As a lognormal input to the generator has a matching power spectrum to its corresponding $N$-body slice, any deviation away from this would be indicative of a generated map. The power spectrum path is constructed such that it calculates an elementwise difference between the lognormal and emulated power spectra. This is then fed into the classification head to aid the critic in differentiating emulated images as any significant deviation from the lognormal power spectrum can be used to easily identify an emulation, thus aiding the critic in achieving its goal. This in turn forces the generator to learn how to maintain the power spectrum of its input so that it can successfully ``fool'' the critic. This information is also able to be backpropagated through the critic and generator networks to ensure that the generator is trained to capture features at all scales.

This approach was favoured over the more traditional method of adding a power spectrum term directly to the loss function as the loss function was found to be  extremely sensitive to the weighting of this additional term, often resulting in non-convergent training. Determining the optimal value of this weighting would require a brute-force hyperparameter search, which would be less efficient than our method of introducing a power spectrum path to the critic and allowing the network to learn the optimal balance between the power spectrum path and the convolutional path. The addition of the power spectrum path introduces $256 \times 256 = 65\,536$ extra parameters to the network, which is computational insignificant in comparison to both the size of the whole \textsc{Psi-GAN} network, as well as the alternative of adding a power spectrum term to the loss function and running a hyperparameter search in order to optimise its weighting.

Both the generator and the critic are trained in tandem in order to minimise the loss function $L_{\mathrm{train}}$, which we choose to be the standard WGAN-GP formulation with an additional term equating to the $l^{2}$ norm between out generated map and the target $N$-body slice:
\begin{equation}
    L_{\mathrm{train}} = L_{\mathrm{G}} + L_{\mathrm{NB}} + L_{\mathrm{GP}} + L_{\mathrm{pixel}},
    \label{eq:loss_train}
\end{equation}
where each component of the training loss is given by:
\begin{equation}
    L_{\mathrm{G}} = \mathbb{E}_{G(\delta_{\mathrm{LN}})}[C(G(\delta_{\mathrm{LN}}))],
    \label{eq:loss_g}
\end{equation}
\begin{equation}
    L_{\mathrm{NB}} = - \mathbb{E}_{\delta_{\mathrm{NB}}}[C(\delta_{\mathrm{NB}})],
    \label{eq:loss_nb}
\end{equation}
\begin{equation}
    L_{\mathrm{GP}} = \lambda_{\mathrm{gp}} \mathbb{E}_{\hat{\delta}} \left[ \left( \lvert \lvert \nabla_{\hat{\delta}} C(\hat{\delta}) \rvert \rvert_{2} -1 \right)^{2} \right],
    \label{eq:loss_gp}
\end{equation}
\begin{equation}
    L_{\mathrm{pixel}} = \lambda_{\mathrm{pixel}} \lvert \lvert G(\delta_{\mathrm{LN}}) - \delta_{\mathrm{NB}} \rvert \rvert^{2}_{2},
    \label{eq:loss_pixel}
\end{equation}
where $G$ and $C$ are the generator and critic networks respectively, $\delta_{\mathrm{LN}}$ and $\delta_{\mathrm{NB}}$ are lognormal and $N$-body simulation slices, $\hat{\delta}$ represents a linear combination of $G(\delta_{\mathrm{LN}})$ and $\delta_{\mathrm{NB}}$\footnote{Specifically, $\hat{\delta} = \alpha \delta_{\mathrm{NB}} + (1-\alpha) G(\delta_{\mathrm{LN}})$ with $\alpha \sim U(0,1)$, where $U(0,1)$ indicates the uniform distribution between 0 and 1. This linear combination means that we constrain the gradient norm to be 1 only along lines that connect real and fake data \citep{wgan_gp}.}, $\mathbb{E}$ represents the expectation over a sample, $\lvert \lvert \cdot \rvert \rvert_{2}$ represents the $l^{2}$ norm, and $\lambda_{\mathrm{gp}}$ and $\lambda_{\mathrm{pixel}}$ are hyperparameters used to control the amount of regularisation from the gradient penalty and $l^{2}$ norm between the generated output and the target. We set $\lambda_{\mathrm{gp}}=10$ and $\lambda_{\mathrm{pixel}}=100$, however we leave the optimisation of these hyperparameters to future work.

The addition of $L_{\mathrm{pixel}}$ to what is otherwise the standard WGAN-GP loss function was motivated by \citet{fast_nbody_sims}, who found this term to aid in generating emulations with accurate statistics but ineffective in producing structure correlated with target $N$-body simulations.

\subsection{Training}
\label{subsec:method_training}

We train using the Adam optimiser \citep{adam} with a base learning rate $\alpha=10^{-4}$, and decay parameters $(\beta_{1}, \beta_{2}) = (0.5, 0.9)$. Following \citet{ttur}, we increase the learning rate for the critic by a factor of $f = 3$ while using the base rate for the generator. We allow the model to train for an initial period of three epochs, after which we half the learning rate after every epoch where the validation loss increases. We also employ gradient clipping to clamp the magnitude of the gradients to a maximum value of $1\,000$. The gradient penalty term in the loss function should act to keep gradients close to unity, however there is a warm up period until it is able to have its intended effect. Clipping the gradients was found to be useful in avoiding overflow errors before the gradient penalty took effect.

We train using a batch size of $6$, and use randomised data augmentation techniques when compiling a batch. The same data augmentations were applied both $\delta_{\mathrm{LN}}$ and $\delta_{\mathrm{NB}}$ and consist of:
\begin{enumerate}
    \item horizontal and vertical flips,
    \item horizontal and vertical translations of $x, y \in [0,512)$ pixels,
    \item rotations of $\theta \in \left\{ 0, \pi/2, \pi, 3\pi/2 \right\}$.
\end{enumerate}
We use $32$-bit floating point precision for numerical stability. A single epoch of training and validation takes ${\sim}15$ hours on a single NVIDIA A100 Tensor Core GPU, and we train for $10$ epochs. Training time is significantly inflated as the critic requires the power spectrum to be calculated for each generated sample in the dataset. However, we accelerate this computation by using a parallelised GPU implementation. We also pre-compute power spectra for all $\delta_{\mathrm{LN}}$ and $\delta_{\mathrm{NB}}$ in our dataset so that they do not need to be calculated during training. Once trained, the generator can process $512$ lognormal slices in ${\sim}2$ minutes on similar hardware.

\subsection{Validation and testing}
\label{subsec:method_val_test}

We validate and test our model using a range of summary statistics which will be described in this section. We save the model after each training epoch, and select the best model using a weighted sum of the absolute percentage error across the summary statistics (excluding the bispectrum and reduced matter bispectrum, due to the complexity of their calculation). We weight the summary statistics such that the power spectrum has a weighting seven times that of the other statistics in order to bias our model selection towards a model that reproduces an accurate power spectrum. We also add redshift-dependent weighting to the validation loss, with redshift $z = 0$ examples being given double the weighting of all other redshifts. This is to bias model selection towards a model that performs well at low redshifts. To test the model, we quantitatively compare these summary statistics for the lognormal slices, generated slices, and $N$-body slices in each of the test sets described in Table~\ref{tab:datasets}.

\subsubsection{Pixel counts histogram}
\label{subsubsec:method_val_test_pixel}

We bin the pixel values of the lognormal, generated, and $N$-body slices into a histogram of $64$ equally sized bins. The ranges that these bins span differ depending on redshift, and were qualitatively chosen in order to ensure that all bins have a count of at least $10$ pixels in order to avoid divide-by-zero errors when computing relative differences. It can be seen in Section~\ref{sec:results} that the lognormal approximation differs significantly from the target $N$-body distribution, while our model aims to improve over the lognormal.

\subsubsection{Peak counts histogram}
\label{subsubsec:method_val_test_peak}

We use peak counts to assess whether the model has learned the non-Gaussian features of the $N$-body field. A peak is defined as a pixel with a higher value than all of its eight surrounding pixels. We bin peak count values into a histogram of $64$ equally sized bins in order to compare non-Gaussian information between different models. Similarly to the pixel counts histogram, the ranges of these bins differ by redshift and were chosen in order to avoid divide-by-zero errors when calculating errors. Peak count statistics have been shown to carry significant cosmological information, especially in cosmic shear studies \citep{peaks_1, peaks_2, peaks_3, peaks_4, peaks_5, peaks_6, peaks_7, peaks_8, peaks_9, peaks_10}.

\subsubsection{Phase difference distribution}
\label{subsubsec:method_val_test_phase}

The phases of Fourier modes are an important measure of non-linearity in the cosmic web. While a Gaussian field exhibits randomised phases, non-linear structure growth introduces correlations into the phases. While the power spectrum relies only on Fourier amplitudes, it has been shown that the phases carry substantial information regarding the structure of the matter overdensity field \citep{phase_information} thus making phase statistics extremely important in analysing the cosmic web.

Many methods exist to quantify phase statistics, including calculating the entropy of Fourier phases and measuring the distribution of phases \cite[see e.g.][]{phase_entropy_1, phase_entropy_2, phase_difference, phase_correlations, fourier_stats}. We focus on the probability distribution of phase differences as described by \citet{phase_difference}, in which the authors define a quantity $D_{k}$ given by:
\begin{equation}
    D_{k} = \Phi_{k+1} - \Phi_{k},
    \label{eq:phase_difference}
\end{equation}
which measures the difference in the phases of adjacent Fourier modes (in a single dimension) $k$ and $k+1$. This can be extended to a two-dimensional field by calculating a set of $D_{k}$ in two orthogonal directions. \citet{phase_difference} find that the distribution of these phase differences $P(D)$ can be described by a von Mises distribution:
\begin{equation}
    P(D) = \frac{1}{2 \pi I_{0}(\kappa)} e^{-\kappa \mathrm{cos}(D - \mu)},
    \label{eq:phase_distribution}
\end{equation}
where $\mu$ is the mean angle which varies from sample to sample, $\kappa$ is a parameter that describes the level of non-linearity, and $I_{0}$ is a modified Bessel function of order zero.

In order to measure $P(D)$ for a dark matter overdensity map, we bin the phase differences into histograms of $64$ equally spaced bins which we use to assess whether the model has correctly learned non-linear growth through phase statistics.

\subsubsection{Power spectrum}
\label{subsubsec:method_val_test_power}

Although the lognormal input to the model and the target $N$-body simulation have the same power spectrum, we cannot ensure that our model does not significantly alter it. In order to assess whether the power spectrum has been significantly changed, we use the estimator
\begin{equation}
    \hat{P}(k) = \frac{1}{N_{\mathrm{modes}}(k)} \sum_{\lvert \mathbf{k} \rvert = k} \lvert \delta(\mathbf{k}) \rvert^{2},
    \label{eq:power_spectrum_estimator}
\end{equation}
where $\delta(\mathbf{k})$ is the Fourier transform of the matter overdensity $\delta(\mathbf{x})$, the summation is performed over all $\mathbf{k}$ vectors with a magnitude of $k$, and $N_{\mathrm{modes}}(k)$ is the number of modes in each $k$ bin.

\subsubsection{Bispectrum}
\label{subsubsec:method_val_test_bi}

Since the power spectrum is unable to capture any information regarding Fourier phases, we can use the matter bispectrum $B(k_{1}, k_{2}, k_{3})$ to quantify non-linear structure. The bispectrum can be seen as a three-point counterpart to the power spectrum \citep{bispectrum}. The bispectrum for a two-dimensional field is defined by the relation:
\begin{equation}
    \langle \delta(\mathbf{k}_{1}) \delta(\mathbf{k}_{2}) \delta(\mathbf{k}_{3}) \rangle = (2 \pi)^{2} \delta_{\mathrm{D}}(\mathbf{k}_{1} + \mathbf{k}_{2} + \mathbf{k}_{3}) B(k_{1}, k_{2}, k_{3}),
    \label{eq:bispectrum}
\end{equation}
where $k_{i}=\lvert \mathbf{k}_{i} \rvert$, all $\mathbf{k}_{i}$ vectors are in the plane of the two-dimensional slice, $\delta_{\mathrm{D}}(\cdot)$ indicates the Dirac delta function, and $\langle \cdot \rangle$ represents an expectation value over all Fourier space.

We also assess the reduced matter bispectrum $Q(k_{1}, k_{2}, k_{3})$ \cite[see e.g.][]{reduced_bispectrum}:
\begin{equation}
    Q(k_{1}, k_{2}, k_{3}) = \frac{B(k_{1}, k_{2}, k_{3})}{P(k_{1})P(k_{2}) + P(k_{1})P(k_{3}) + P(k_{2})P(k_{3})}.
    \label{eq:reduced_bispectrum}
\end{equation}

We measure the bispectra and reduced matter bispectra based on an estimator of the binned bispectrum \citep{piinthesky_flatsky, piinthesky_fullsky}. Bispectra can be measured along different triangle configurations, and it is important to consider many configurations when using the bispectrum as a statistical tool in order to break degeneracies when inferring cosmological parameters \citep{bispectrum_degen}. Therefore we measure the bispectra and reduced bispectra using multiple configurations of $(k_{1}, k_{2}) \in \{0.05, 0.2, 0.4, 0.6\}~h~\mathrm{Mpc}^{-1}$, which span both regular configurations and squeezed bispectra configurations. In Section~\ref{sec:results}, we report \textsc{Psi-GAN}'s performance for only two of these configurations: $(k_{1}, k_{2}) = (0.4, 0.4)~h~\mathrm{Mpc}^{-1}$ and $(k_{1}, k_{2}) = (0.4, 0.6)~h~\mathrm{Mpc}^{-1}$, however our full set of tests show that \textsc{Psi-GAN}'s performance is similar over all configurations tested.

\section{Results}
\label{sec:results}

\begin{figure*}
    \centering
    \includegraphics[width=0.7\textwidth]{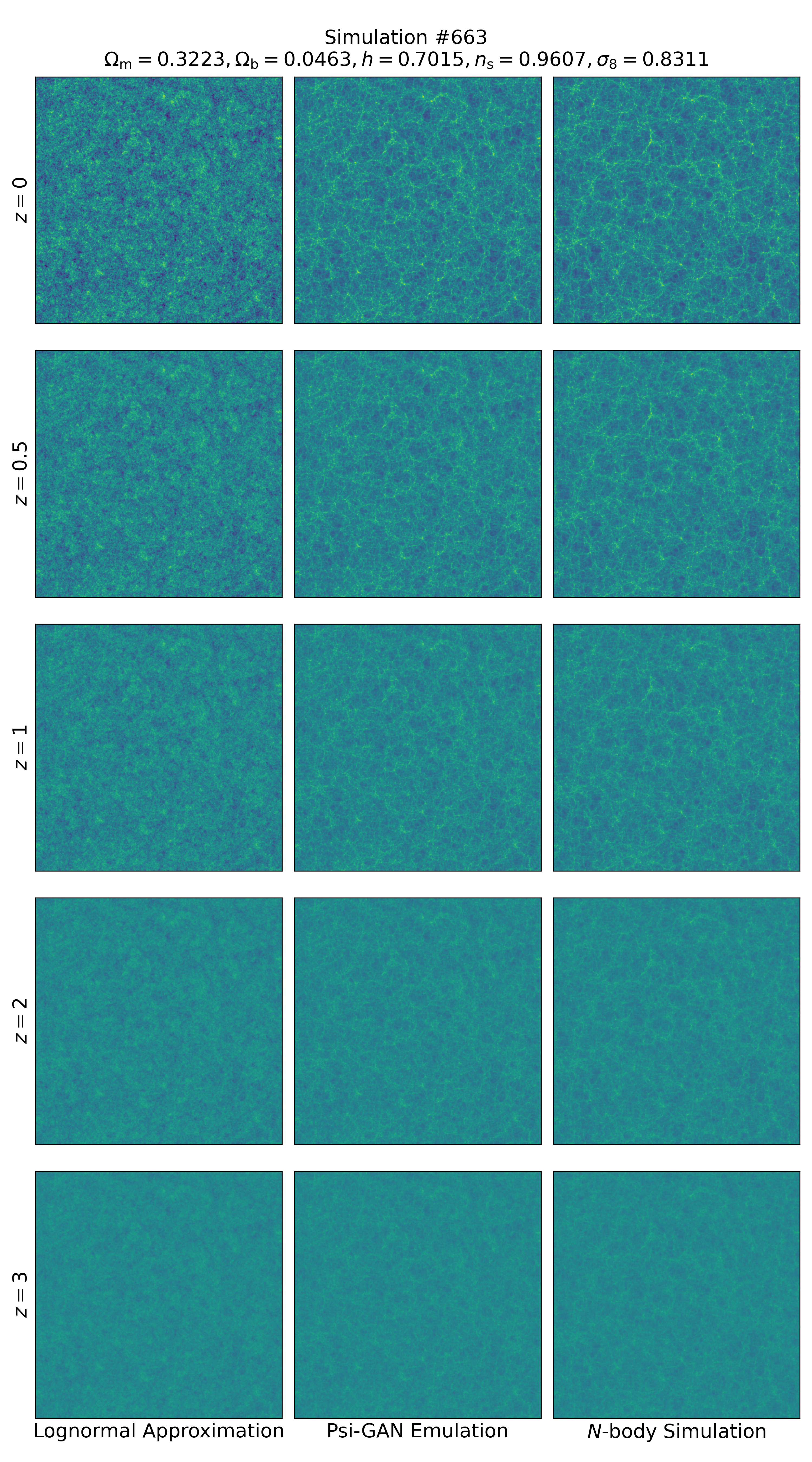}
    \caption{A set of examples for our ``fiducial'' cosmology, showing lognormal random fields (left), \textsc{Psi-GAN} generated emulations (centre), and $N$-body simulations (right) for all redshift bins.}
    \label{fig:examples_663}
\end{figure*}

Visual inspection shows that \textsc{Psi-GAN} is able to accurately reproduce the structure of the cosmic web across all redshift bins. Figure~\ref{fig:examples_663} shows a set of examples for simulation $\#663$, our ``fiducial'' cosmology.

\begin{figure*}
    \centering
    \includegraphics[width=0.7\textwidth]{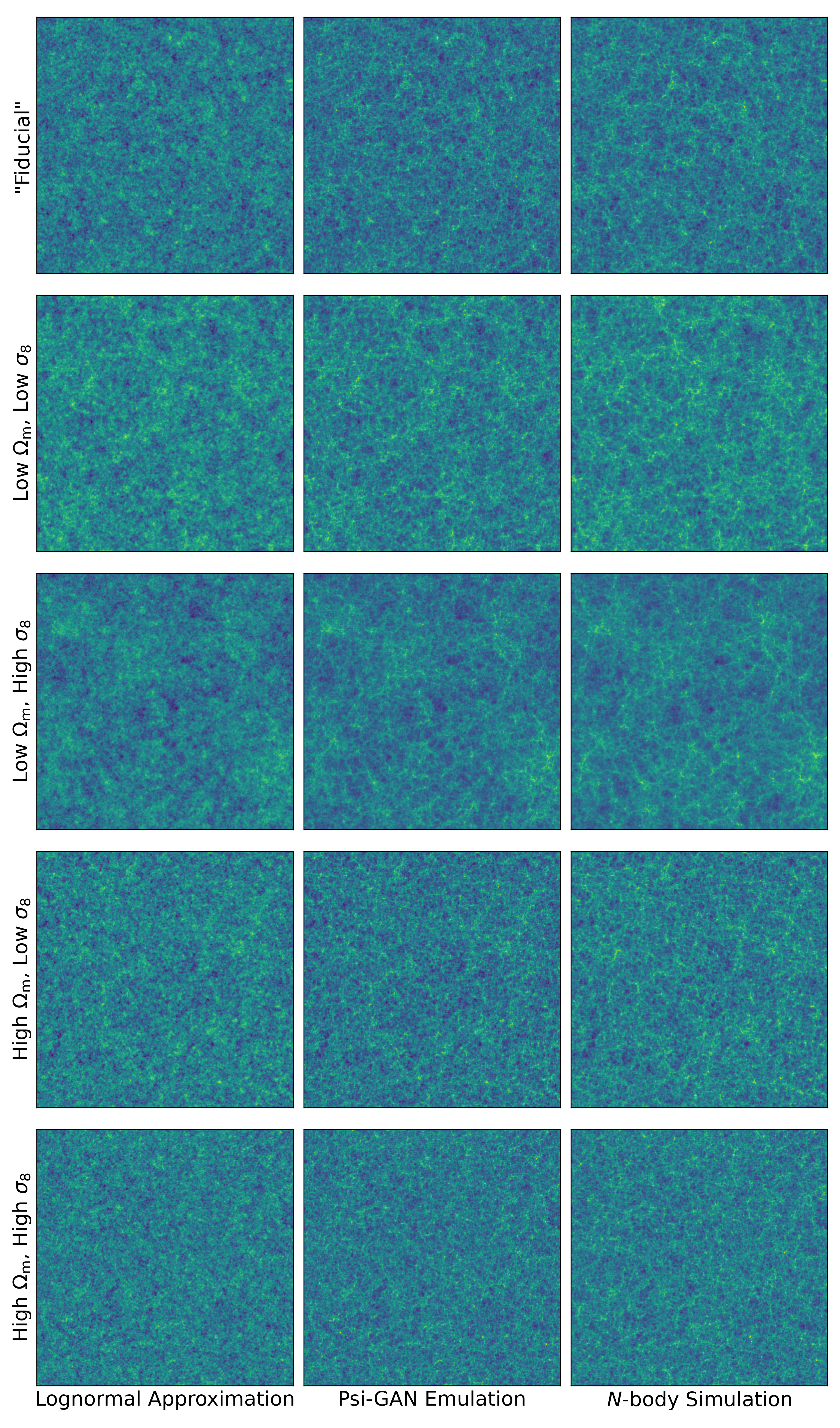}
    \caption{A set of examples for our ``fiducial'' cosmology and extreme values of the $\Omega_{\mathrm{m}}, \sigma_{8}$ subspace (as defined in Table~\ref{tab:example_cosmo}), showing lognormal random fields (left), \textsc{Psi-GAN} generated emulations (centre), and $N$-body simulations (right) for redshift $z = 0$.}
    \label{fig:examples_cosmo}
\end{figure*}

In addition, Figure~\ref{fig:examples_cosmo} shows example maps at redshift $z = 0$ for our ``fiducial'' cosmology, as well as extreme values of the $\Omega_{\mathrm{m}}, \sigma_{8}$ subspace. Table~\ref{tab:example_cosmo} shows the values for the chosen cosmologies.

\begin{table}
    \centering
    \caption{The cosmologies used to demonstrate \textsc{Psi-GAN}'s emulations in Figure~\ref{fig:examples_cosmo}.}
    \label{tab:example_cosmo}
    \begin{tabular}{cccccc}
        \hline
        Cosmology                                       & $\Omega_{\mathrm{m}}$ & $\Omega_{\mathrm{b}}$ & $h$       & $n_{\mathrm{s}}$  & $\sigma_{8}$  \\
        \hline
        ``Fiducial''                                    & $0.3223$              & $0.04630$             & $0.7015$  & $0.9607$          & $0.8311$      \\
        Low $\Omega_{\mathrm{m}}$, Low $\sigma_{8}$     & $0.1663$              & $0.04783$             & $0.6173$  & $1.1467$          & $0.6461$      \\
        Low $\Omega_{\mathrm{m}}$, High $\sigma_{8}$    & $0.1289$              & $0.06325$             & $0.7293$  & $1.1537$          & $0.9489$      \\
        High $\Omega_{\mathrm{m}}$, Low $\sigma_{8}$    & $0.4599$              & $0.04055$             & $0.7287$  & $0.8505$          & $0.7011$      \\
        High $\Omega_{\mathrm{m}}$, High $\sigma_{8}$   & $0.4423$              & $0.03533$             & $0.8267$  & $1.0009$          & $0.9151$      \\
        \hline
    \end{tabular}
\end{table}

Although \textsc{Psi-GAN} was trained with the goal of reproducing accurate statistics, we also see some correlations in structure between \textsc{Psi-GAN} emulations and $N$-body simulations. This emerges as the GAN framework aims to reproduce maps matching $N$-body simulations starting from correlated lognormal maps. However, for the applications we are interested in, we mainly care about summary statistics, and therefore choose to assess the model's performance by how well it is able to reproduce those, as opposed to assessing any apparent structure correlation.

\subsection{Randomised test set}
\label{subsec:results_test_set}

\begin{figure*}
    \centering
    \includegraphics[width=0.9\textwidth]{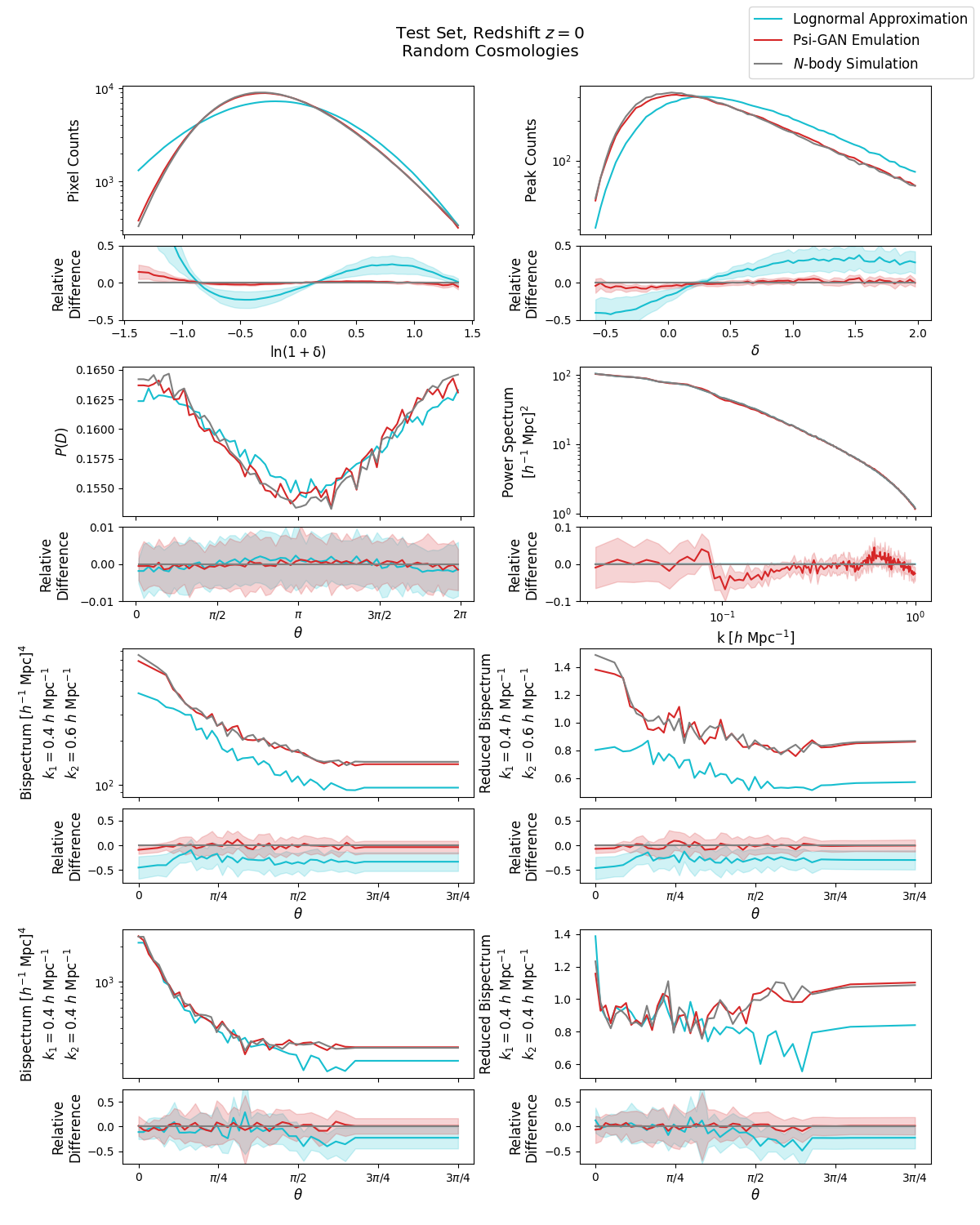}
    \caption{A comparison of the statistical tests as described in Section~\ref{subsec:method_val_test} for the lognormal approximation (cyan), $N$-body simulation (grey), and \textsc{Psi-GAN} emulation (red) on the randomised test set. The metrics displayed are as follows (from top to bottom, and left to right): pixel counts, peak counts, phase difference distribution, power spectrum, bispectrum with $(k_{1}, k_{2}) = (0.4, 0.6)~h~\mathrm{Mpc}^{-1}$, reduced bispectrum with $(k_{1}, k_{2}) = (0.4, 0.6)~h~\mathrm{Mpc}^{-1}$, bispectrum with $(k_{1}, k_{2}) = (0.4, 0.4)~h~\mathrm{Mpc}^{-1}$, and reduced bispectrum with $(k_{1}, k_{2}) = (0.4, 0.4)~h~\mathrm{Mpc}^{-1}$. The relative performance with respect to the $N$-body simulation can be seen in in the bottom panel for each test, along with the respective uncertainties. We show that \textsc{Psi-GAN} outperforms the lognormal approximation across all tests with the exception of the power spectrum, in which we see small discrepancies within $5$ per cent.}
    \label{fig:test_set_z=0}
\end{figure*}

Figure~\ref{fig:test_set_z=0} shows the results of all eight test metrics for our randomised test set for redshift $z = 0$. On the top panel for each metric we show the mean value averaged over $64$ examples of the $N$-body simulation, the \textsc{Psi-GAN} emulation, and the lognormal approximation. On the bottom panel we show the relative difference with respect to the $N$-body simulation for each model. We include uncertainties only on the bottom panel for the sake of visual clarity.

\begin{figure*}
    \centering
    \includegraphics[width=0.9\textwidth]{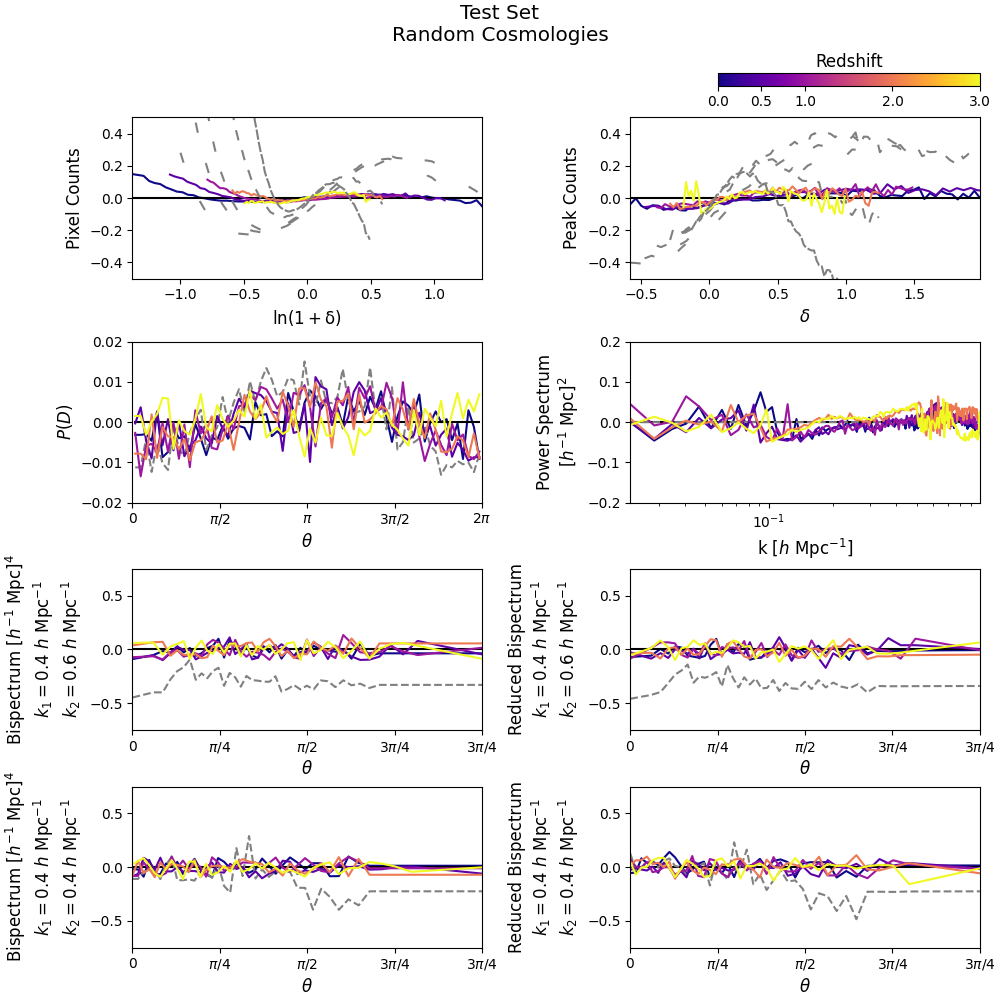}
    \caption{A comparison of the relative differences in the statistical tests as described in Section~\ref{subsec:method_val_test} for the lognormal approximation (grey dashed line with looser dash spacing indicating a lower redshift snapshot), $N$-body simulation (black line), and \textsc{Psi-GAN} emulation (lines coloured with respect to the redshift colour at the top of the figure) on the randomised test set. The metrics shown are the same as in Figure~\ref{fig:test_set_z=0}. We show that \textsc{Psi-GAN} outperforms the lognormal approximation across all tests with the exception of the power spectrum, in which we see small discrepancies within $5$ per cent. Uncertainties are similar to those shown in Figure~\ref{fig:test_set_z=0}, but are omitted here for visual clarity.}
    \label{fig:test_set_differences}
\end{figure*}

In addition, in Figure~\ref{fig:test_set_differences}, we show the relative differences averaged over $64$ examples for each model, for all redshifts when compared to $N$-body simulations. We also display the relative differences for the lognormal approximation for comparison. We show all redshift snapshots on the top two panels, however we only show redshift $z = 0$ on the remaining panels for visual clarity.

\textsc{Psi-GAN} shows an improvement over the lognormal approximation with the sole exception of the power spectrum. The lognormal approximation was designed to have an identical power spectrum to the $N$-body simulation, so this was an expected result. However, we can say that the power spectrum path in the critic of \textsc{Psi-GAN} was effective in constraining the power spectrum so that it was not altered by more than ${\sim}5$ per cent. Initial trials of an GAN using a fully convolutional critic (i.e. without the power spectrum path) saw differences in the power spectrum between the emulation and the $N$-body simulation of ${\sim}20$ per cent. Thus we can be confident that our critic architecture is effective in maintaining the power when transforming a lognormal random field.

We see agreement to within ${\sim}5$ per cent for all metrics, with the exception of the pixel counts at low values of $\mathrm{ln}(1 + \delta)$. This is due to the baseline count for the $N$-body simulation being very low (${\sim}10^{2}$), and thus making the relative differences sensitive to small changes in pixel counts. We also believe that this issue is partially caused by the model's architecture more easily modelling higher values of $\mathrm{ln}(1 + \delta)$ due to the use of the GELU activation function \citep{gelu}, which is more expressive at positive values.

\subsection{Redshift interpolation}
\label{subsec:results_redshift}

\begin{figure*}
    \centering
    \includegraphics[width=0.9\textwidth]{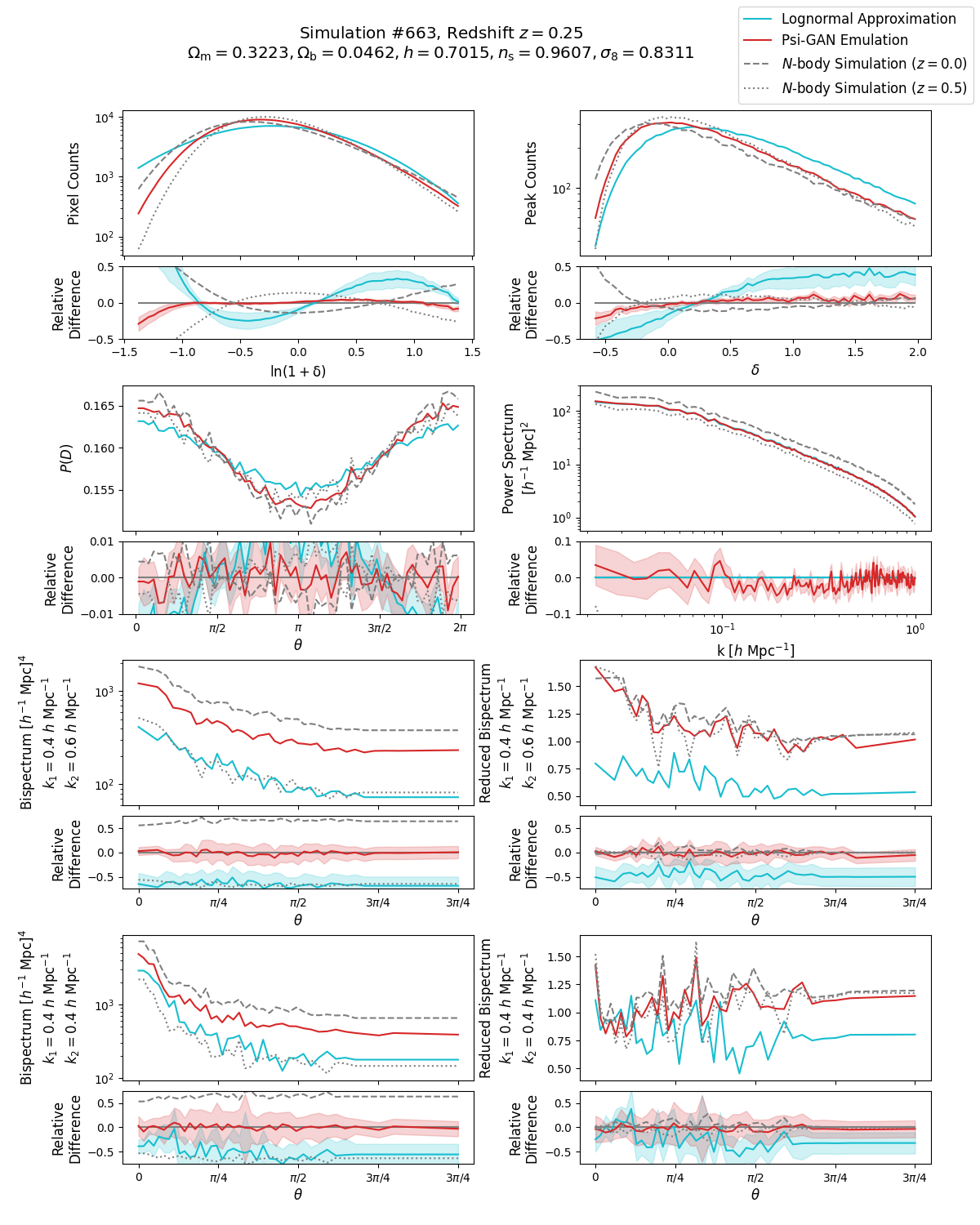}
    \caption{Similar to Figure~\ref{fig:test_set_z=0}, but for the test set for interpolating redshift at $z = 0.25$. As we have no $N$-body snapshot to act as the ground truth, relative differences are displayed with respect to a value interpolated between the two adjacent redshift snapshots. The dashed grey lines show the measured values for these snapshots. Also, please note that the grey dashed lines line outside of the range of the y-axis for the relative differences plot of the power spectrum.}
    \label{fig:interpolate_z_z=0.25}
\end{figure*}

Figure~\ref{fig:interpolate_z_z=0.25} displays similar results to Figure~\ref{fig:test_set_z=0}, but for our redshift interpolation test at $z = 0.25$. On the bottom panel we show the relative difference with respect to a value interpolated between the two adjacent redshift snapshots ($z = 0$ and $z = 0.5$) as we have no $N$-body snapshot to act as the ground truth.

It can be seen that \textsc{Psi-GAN} improves on the lognormal approximation across all metrics. Although not much can be quantitatively said about the performance of \textsc{Psi-GAN} with respect to the $N$-body snapshots, we can qualitatively say that the results lie reasonably between the upper and lower bounds set by the adjacent redshift snapshots (as shown in Figure~\ref{fig:interpolate_z_z=0.25}), and within ${\sim}5$ per cent of an interpolated baseline. We can also see that \textsc{Psi-GAN}'s metrics intercept the $N$-body snapshots exactly at cross-over points for the pixel counts, peak counts, and phase difference distributions. We can also see that the power spectrum does not vary by more than $3$ per cent from the lognormal approximation at redshift $z = 0.25$, again showing the effectiveness of the power spectrum path in \textsc{Psi-GAN}'s critic.

Our second redshift interpolation test at redshift $z = 0.75$ showed similar results to the test at $z = 0.25$, but are not shown here for brevity. All metrics showed agreement with $N$-body simulations to within ${\sim}5$ per cent, with the power spectrum showing closer agreement to ${\sim}3$ per cent. The only case of the agreement differing by more than this when the pixel and peak counts histograms were at a very low baseline value (${\sim}10^{2}$), where we saw discrepancies of ${\sim}15$ per cent.

\subsection{Cosmology interpolation}
\label{subsec:results_cosmology}

\begin{figure*}
    \centering
    \includegraphics[width=0.9\textwidth]{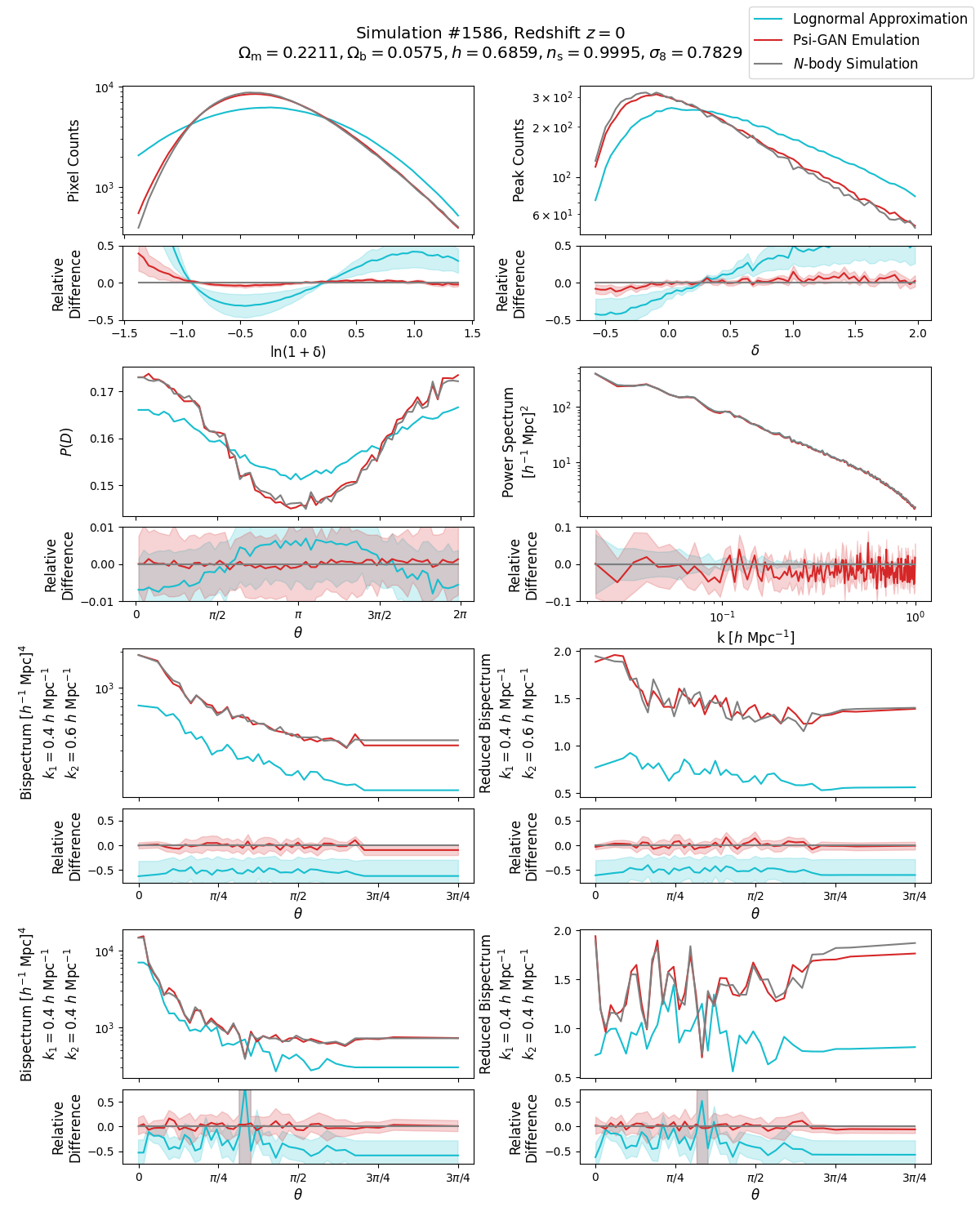}
    \caption{Similar to Figure~\ref{fig:test_set_z=0}, but for the test set for interpolating cosmology on simulation $\#1586$ at redshift $z = 0$.}
    \label{fig:interpolate_cosmo_1586_z=0}
\end{figure*}

Figure~\ref{fig:interpolate_cosmo_1586_z=0} displays similar results to Figure~\ref{fig:test_set_z=0}, but for our cosmology interpolation test for simulation $\#1586$ at redshift $z = 0$.

\begin{figure*}
    \centering
    \includegraphics[width=0.9\textwidth]{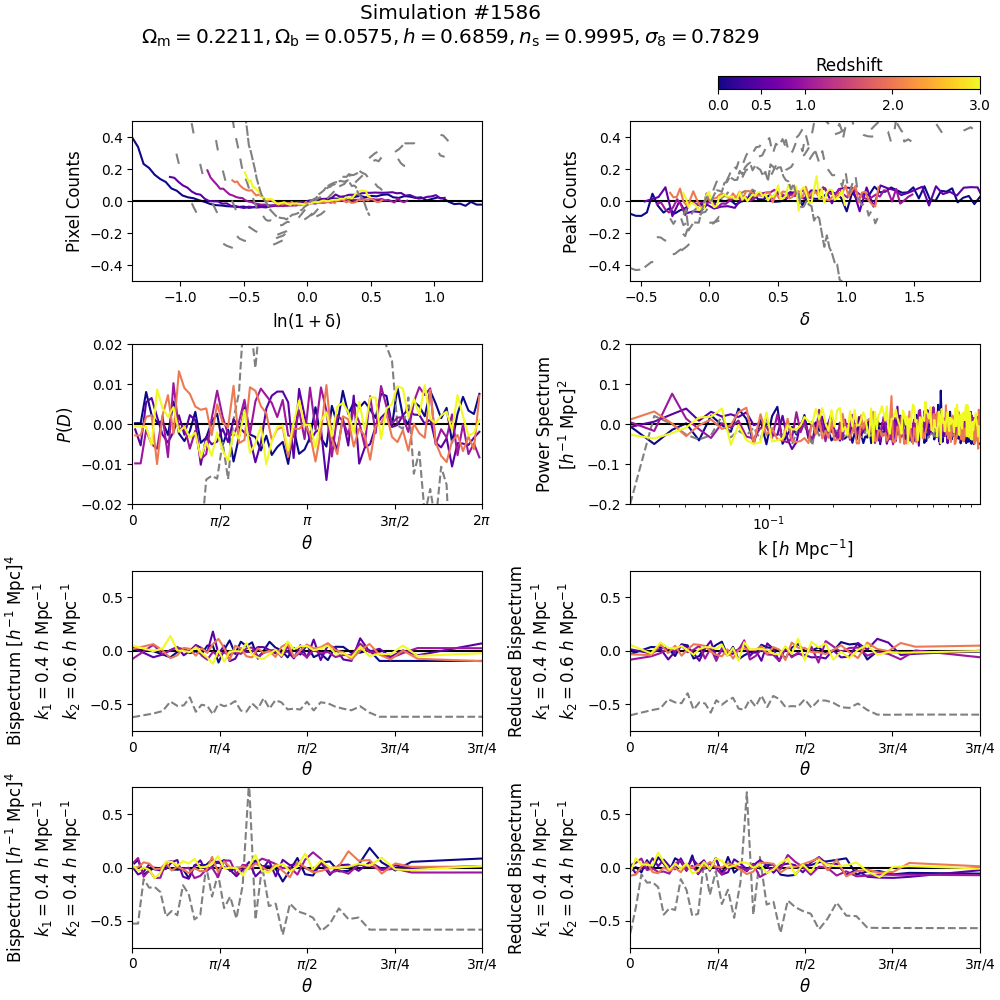}
    \caption{Similar to Figure~\ref{fig:test_set_differences}, but for the test set for interpolating cosmology on simulation $\#1586$.}
    \label{fig:interpolate_cosmo_1586_differences}
\end{figure*}

Figure~\ref{fig:interpolate_cosmo_1586_differences} displays the relative differences for all redshifts tested (similar to Figure~\ref{fig:test_set_differences}). Our second cosmology interpolation test for cosmology $\#815$ showed similar results to those shown for simulation $\#1586$.

\textsc{Psi-GAN} shows an improvement over the lognormal approximation, again with the sole exception of the power spectrum which was constrained so that it was not altered by more than ${\sim}5$ per cent. We do see greater discrepancies in the power spectrum compared to the previous tests. We believe that this discrepancy can be explained by the node coverage over cosmology-space when compared to redshift-space.

Redshift is a one-dimensional space which we cover with $5$ nodes at snapshots of $z \in \{ 0, 0.5, 1, 2, 3 \}$. However, cosmology is a five-dimensional space (i.e. we condition on five cosmological parameters) which we cover with $2\,000$ nodes. In order to cover cosmology-space with the same density as we cover redshift-space, we would require $5^{5} = 3\,125$ nodes in cosmology space. We are significantly short of this number, requiring $56.25$ per cent more simulations than are part of the Latin hypercube suite.

\subsection{Model analysis through saliency mapping}
\label{subsec:results_model}

Saliency mapping is a field of techniques used to produce visual explanations of the behaviour of computer vision models \citep{smoothgrad}. These explanations take the form of heatmaps which aim to highlight which areas are most important for the model to reach a specific output. These scores are often computed by taking gradients of the output in question with regards to the input image \cite[see e.g.][for an overview of various methods used in computer vision]{saliency_compare, saliency_benchmark}.

Saliency mapping has been explored in astrophysics through a variety of applications such as measuring galaxy bar lengths from morphology classification models \citep{expgal}, and qualitatively investigating model behaviour for both AGN classification models \citep{agn_saliency} and cosmological parameter estimation models \citep{deeplss}.

In order to investigate potential model improvements, we perform saliency mapping on the output of the critic with respect to a \textsc{Psi-GAN} emulation with the hope of discovering any features that may be tell-tale signs of a certain map being an emulation. We use \textsc{SmoothGrad-Squared} \citep{saliency_benchmark} to visualise which areas of an emulation are used by the critic to identify it as an emulation as opposed to an $N$-body simulation.

\textsc{SmoothGrad-Squared} extends vanilla saliency \citep{saliency}, in which the saliency map $L^{c}$ is created by simply taking the gradient of the output with regards to each input pixel. Vanilla saliency has been shown to be unstable \citep{saliency_compare} due to gradients exhibiting large fluctuations with respect to pixel values, which creates excess noise in the resultant saliency maps. \textsc{SmoothGrad-Squared} aims to improve this limitation by creating $N_{\mathrm{sg}}$ visually similar samples of each image by adding a small amount of Gaussian noise to the original to create each sample, calculating a saliency map for each sample, and then aggregating these to produce a final saliency map:
\begin{equation}
    \hat{L}^{c} (x) = \frac{1}{N_{\mathrm{sg}}} \sum^{N_{\mathrm{sg}}}_{i=1} L^{c} [x + \mathcal{N}(0, \sigma^{2})]^{2},
    \label{psigan_eq:sgrad_2}
\end{equation}
where $\mathcal{N}(0, \sigma^{2})$ is the probability density function for a Gaussian distribution with a mean of $0$ and a standard deviation of $\sigma^{2}$. Here we adopt the notation used by \citet{saliency_benchmark} in which $L^{c}$ is a vanilla saliency map, and $\hat{L}^{c}$ is the \textsc{SmoothGrad-Squared} saliency map that results from the squaring and aggregation of the saliency maps for each sample $x + \mathcal{N}(0, \sigma^{2})$. Throughout this section we use values of $N_{\mathrm{sg}} = 256$ and $\sigma = 0.2$ to control the number of samples, and the Gaussian noise used in the \textsc{SmoothGrad-Squared} algorithm, respectively.

\begin{figure*}
    \centering
    \includegraphics[width=\textwidth]{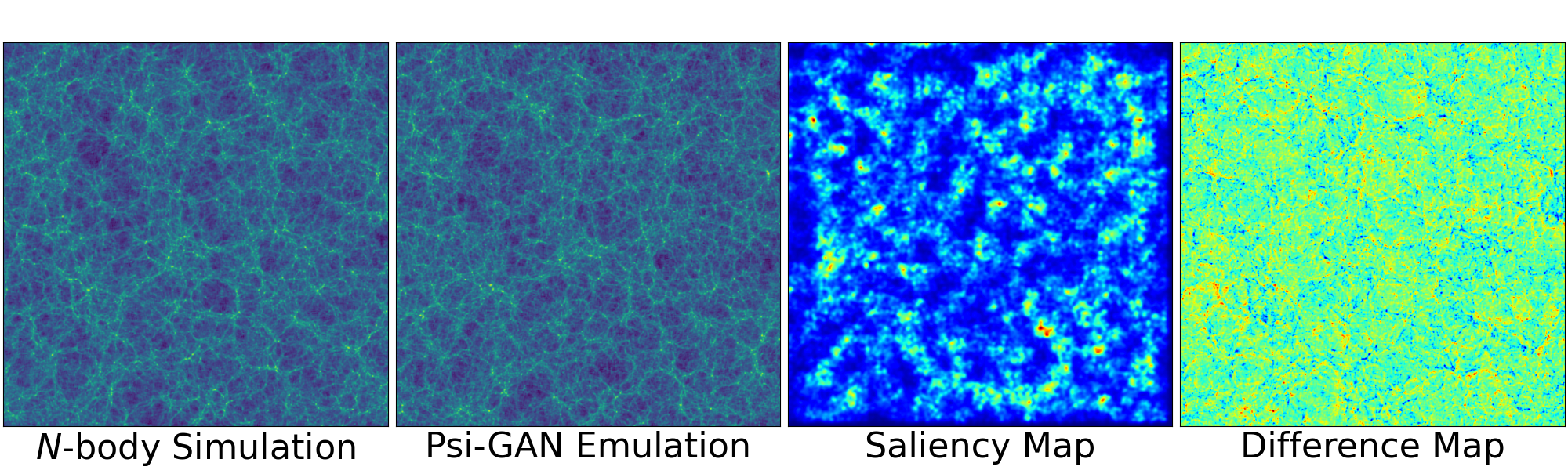}
    \caption{An example showing the the dark matter distribution field from an $N$-body simulation (left) and an emulation generated by \textsc{Psi-GAN} (centre left). We use \textsc{SmoothGrad-Squared} to produce a saliency map (centre right) visualising which parts of the emulation were most important for the critic when determining whether the emulation was real or fake. We also visualise the difference map (right), showing the differences between the $N$-body simulation and the \textsc{Psi-GAN} emulation.}
    \label{fig:xai_critic}
\end{figure*}

Figure~\ref{fig:xai_critic} shows an example at $z = 0$ of an $N$-body simulation, a corresponding emulation produced by \textsc{Psi-GAN}, as well as the \textsc{SmoothGrad-Squared} saliency map, and a difference map.

We examined many such examples in order to visually identify any salient features that are highlighted in the saliency maps. However, we were unable to find any visual correlation between the saliency map and the other visualised maps. We assumed that this must be because the critic uses extremely small-scale features, or long-range correlations (which the human eye is poor at identifying) in order to differentiate \textsc{Psi-GAN} emulations and $N$-body simulations.

\begin{figure}
    \centering
    \includegraphics[width=\columnwidth]{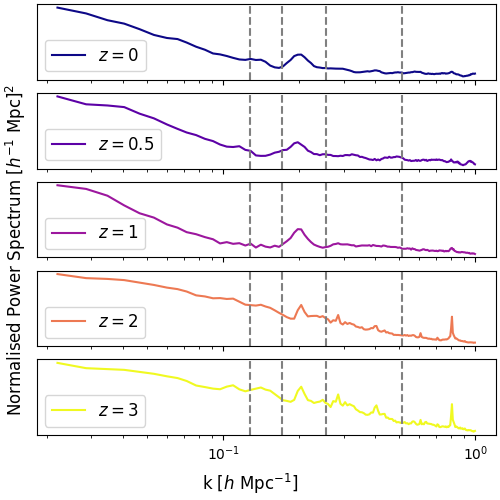}
    \caption{The normalised power spectra of \textsc{SmoothGrad-Squared} saliency maps evaluated on $64$ examples for each redshift snapshot for our ``fiducial'' cosmology. Vertical, dashed-grey lines indicate wavenumbers of integer multiples of the pixel width.}
    \label{fig:xai_power_spectrum}
\end{figure}

We also measure the power spectra of the saliency maps in order to investigate which scales are the limiting factor in \textsc{Psi-GAN}'s emulations. Figure~\ref{fig:xai_power_spectrum} shows the power spectra for each redshift bin averaged over $64$ example maps and normalised such that the maximum value for each redshift is equal to $1$.

Taking the power spectra of the saliency maps shows us that \textsc{Psi-GAN} performs well over all scales. The power spectra show that long-range correlations are slightly more present in the saliency maps when compared to small-scale features. This indicates that \textsc{Psi-GAN} struggles to capture large scales in comparison to small scales, and that long-range correlations are the limiting factor in our architecture's ability to accurately emulate the cosmic web.

We also see peaks at ${\sim}0.2~h~\mathrm{Mpc}^{-1}$ corresponding to a value of $2.5$ times the pixel width. This indicates that \textsc{Psi-GAN} exhibits small amounts of artefacting at small-scales. Although we do not have the computational resources to fully diagnose the cause of this, we believe that it is likely due to the interaction between the scale factor $2$ up-sampling and down-sampling used in the architecture, and the $9 \times 9$ convolutional filter used in the ConvNeXt blocks (see Appendix~\ref{app:model} for further details). The convolutional filter propagates information from $4.5$ pixels away from the centre point of the central pixel, leading to the half-integer pixel width artefacting.

For redshifts $z = 2$ and $z = 3$ we also see a sharp peak at ${\sim}0.8~h~\mathrm{Mpc}^{-1}$. However, as this is on the sub-pixel scale we have no control over it, and we believe that its presence is due to the interpolation algorithm used by \textsc{nbodykit} when measuring the power spectrum.

\section{Conclusions}
\label{sec:conc}

In this paper, we used the Quijote simulations to train a machine learning model (\textsc{Psi-GAN}) capable of transforming two-dimensional flat-sky lognormal random fields of the dark matter overdensity field into more realistic samples across a continuous redshift and cosmology space. \textsc{Psi-GAN} takes the form of a generative adversarial network, with a U-Net generator and a novel critic which uses the power spectrum of the generated samples as a means for discrimination.

We have extensively tested \textsc{Psi-GAN} in a broad series of tests covering: the model's training domain across all redshift ranges, the model's ability to interpolate between the given redshift bins, and the model's ability to interpolate between cosmologies at all redshifts. We observe that \textsc{Psi-GAN} has a closer agreement with $N$-body simulations when compared to the lognormal approximation across statistical tests that probe non-Gaussian features (such as peak counts, phase statistics, and bispectra). \textsc{Psi-GAN} is able to reproduce the bispectra and peak count distributions of $N$-body simulations to ${\sim}5$ per cent, while the lognormal approximation displays a discrepancy of ${\sim}25$ per cent. Due to our novel critic architecture, \textsc{Psi-GAN} is also able to match the power spectrum of the target $N$-body simulation, with relative differences of ${\sim}5$ per cent.

The largest shortcoming of \textsc{Psi-GAN} is its slightly weaker performance in constraining the power spectrum when tasked with interpolating between cosmologies. In our tests, the power spectrum of samples generated by \textsc{Psi-GAN} showed less agreement with $N$-body simulations when interpolating between the cosmologies used in the Quijote simulations. An approximately ${\sim}50$ per cent greater coverage of cosmology space should be enough to reduce this shortcoming, however this would require significant resources to generate. Another potential method of improving this would be to pre-train the model architecture to be able to reconstruct lognormal random fields across an extensive dataset before fine-tuning the model to transform lognormal random field to more accurate emulations.

We used saliency mapping techniques to investigate further architectural improvements to \textsc{Psi-GAN}, which highlighted a slight weakness in capturing long-range correlations as well as a small issue with pixel-scale artefacting. Increasing the depth of the generator by another step (i.e. including an extra set of down-sample and up-sample blocks) should help \textsc{Psi-GAN} model long-range correlations better as the latent space will be more compressed and information will propagate more efficiently across the simulation box. Adding additional ConvNeXt blocks after the first convolution, and before the last convolution should aid in modelling small scales and reduce artefacting, as well as reduce the asymmetry between modelling negative and positive values of the matter density field as discussed in Section~\ref{subsec:results_test_set}.

Another architectural change that could improve performance is to replace the pre-trained ResNet-50 model in the critic with a more powerful option, such as the EfficientNet \citep{efficientnet} or RegNet \citep{regnet} architectures. However, all of these architectural changes will lead to a significant increase in training time which would require state-of-the-art hardware (although inference time should remain unchanged).

To meet our long-term goal of building a full-sky emulator to integrate into \textsc{Glass}, we will have to extend our work to the sphere. The Gower Street simulation suite \cite[currently consisting of $791$ full-sky $N$-body simulations with varying cosmology;][]{gower_street_simulations} provides us with a dataset for training, however it is not as extensive as the Quijote simulation suite used in this paper. Nevertheless, we see two potential avenues for future work on this problem: graph neural networks \cite[see e.g.][for an example pertaining to meteorology]{graphcast}, and rotationally-equivariant convolutions on the sphere \cite[see e.g.][]{disco, gansky}.

\section*{Carbon Intensity Statement}

All work that went into this paper was tracked via ``Weights and Biases'',\footnote{wandb.ai} which allows us calculate that we used a total of ${\sim}2\,000$ GPU hours throughout this work. The majority of this was on NVIDIA A100 Tensor Core GPUs, which had a time-averaged power consumption rate of ${\sim}0.19~\mathrm{kW}$ ($0.2~\mathrm{kW}$ during training and $0.1~\mathrm{kW}$ during validation and testing), thus resulting in a total power consumption of ${\sim}380~\mathrm{kW~h}$.

Using the average carbon intensity of the UK power grid in 2024 \cite[measured at ${\sim}120~\mathrm{gCO_{2}eq}~\mathrm{kW^{-1}~h^{-1}}$;][]{carbon_intensity}, we estimate that we have emitted a total of ${\sim}45.6~\mathrm{kgCO_{2}eq}$ as the result of this work, roughly equivalent to that of a driving from London to Edinburgh and back (${\sim}1\,300~\mathrm{km}$) in a plug-in hybrid car.

We have removed the carbon emissions emitted due to this project from the atmosphere via the funding of carbon capture schemes through the Wren Trailblazer Portfolio.\footnote{www.wren.co/}

\section*{Acknowledgements}

We would like thank William Coulton for the use of their \textsc{PiInTheSky} code, which was used for bispectrum estimation.

PB is supported by the STFC UCL Centre for Doctoral Training in Data Intensive Science. BJ acknowledges support by the ERC-selected UKRI Frontier Research Grant EP/Y03015X/1 and by STFC Consolidated Grant ST/V000780/1. OL acknowledges STFC Consolidated Grant ST/R000476/1 and visits to All Souls College and the Physics Department, University of Oxford. DP was supported by a Swiss National Science Foundation (SNSF) Professorship grant (No. 202671), and by the SNF Sinergia grant CRSII5-193826 ``AstroSignals: A New Window on the Universe, with the New Generation of Large Radio-Astronomy Facilities''.

This work used computing equipment funded by the Research Capital Investment Fund (RCIF) provided by UKRI, and partially funded by the UCL Cosmoparticle Initiative.

\section*{Data Availability}

All code required to reproduce this work has been made publicly available. The code includes a readme detailing the steps required to reproduce the study, including downloading all data and setting the seeds used when randomly splitting the datasets and augmenting training data. \href{https://github.com/prabhbhambra13/psi-gan/}{GitHub \faGithub}.




\bibliographystyle{mnras}
\bibliography{bibliography} 




\appendix

\section{Model architecture}
\label{app:model}

\begin{figure*}
    \centering
    \includegraphics[width=0.75\textwidth]{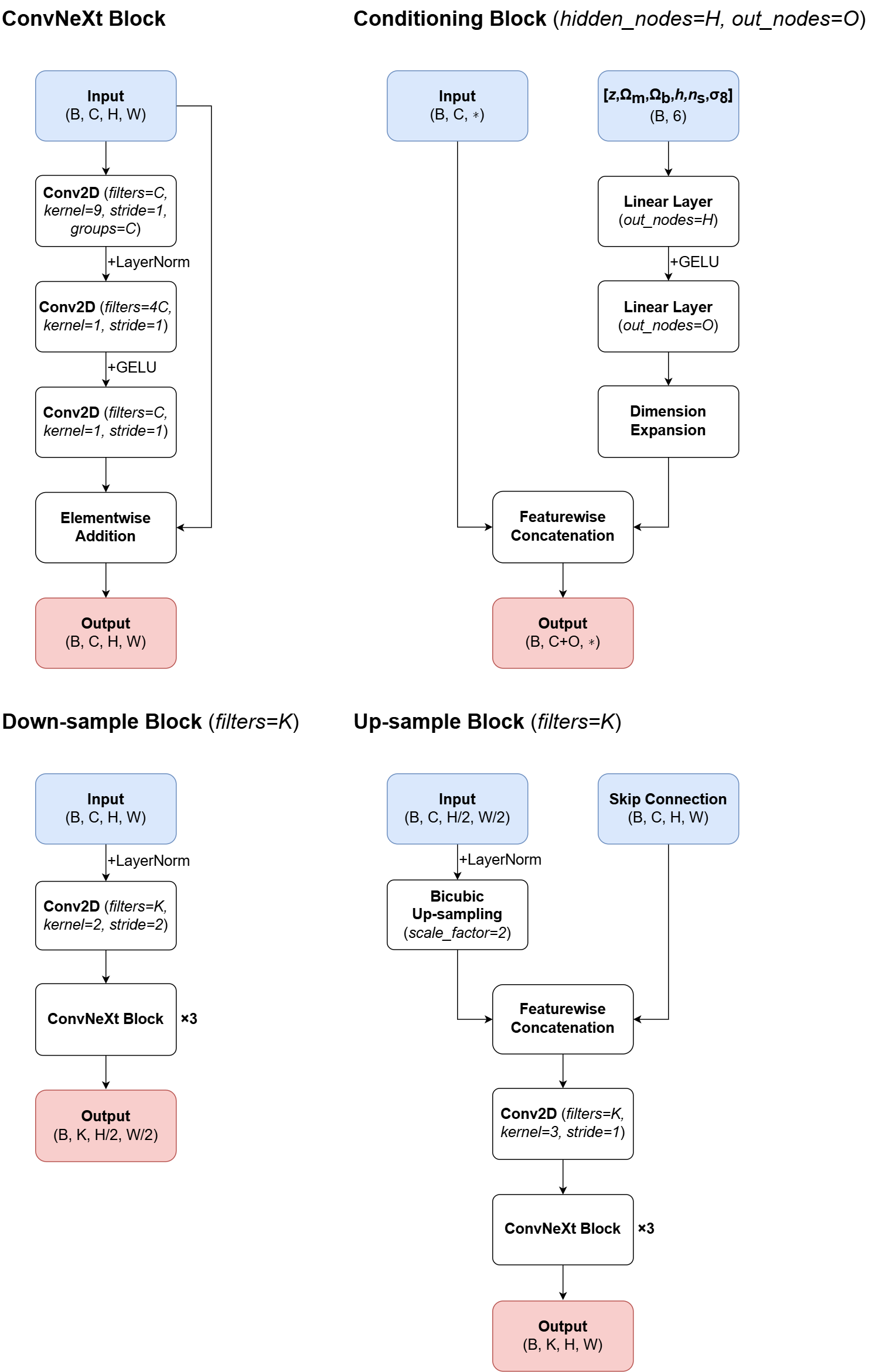}
    \caption{Schematics showing the four building blocks used in the construction of our generator and critic. We also provide information, in parentheses, regarding the dimensions for both the inputs (blue) and outputs (red) in roman font, and the hyperparameters of each layer and block in italics. Dimensions are quoted in the ``batch, channels, $\ast$'' convention, where $\ast$ represents any number of latent dimensions and (B, C, H, W) is used as a placeholder for the dimensions of two-dimensional inputs and outputs (with any lower-dimensional inputs and outputs following a similar convention). All convolutional layers use circular padding in order to maintain the height and width of the input.}
    \label{fig:blocks}
\end{figure*}

Our generator consists of a ConvNeXt-inspired, conditional U-Net \citep{unet}, constructed from the four types of computational blocks shown in Figure~\ref{fig:blocks}. Our ConvNeXt block \citep{convnext} consists of a depthwise separable $9 \times 9$ convolution followed by two $1 \times 1$ convolutions, as well as a residual connection. This architecture aims to efficiently process the input and share information across long ranges and is used as the main processing block for the generator. The conditioning block is used to inject information regarding the redshift and cosmology into the network. This is done by simply taking the $6$ conditioning labels ($z, \Omega_{\mathrm{m}}, \Omega_{\mathrm{b}}, h, n_{\mathrm{s}}, \sigma_{8}$) and embedding them through a two-layered multi-level perceptron. We then expand the dimensions of the embeddings to match that of the input, before finally concatenating this with the input. The conditioning block has hyperparameters controlling the number of hidden and output nodes in the embedding network, which can be used to compress or expand the dimensions of the embedded labels. Following the ConvNeXt architecture, we have separated down-sampling and up-sampling operations away from the main computational block. The down-sample block consists of down-sampling the input using a $2 \times 2$ convolution with a stride of $2$, and then processing the result with three sequential ConvNeXt blocks. The up-sample block takes two inputs, one from the previous step in the generator and another from a skip connection. The first input is up-sampled via bicubic interpolation with a scale-factor of $2$ to match the dimensions of the skip connection. These are then concatenated before being processed through a $3 \times 3$ convolution, followed by three ConvNeXt blocks. Both the down-sample and up-sample blocks have a single parameter that controls the number of filters used in the initial convolution in each block. This is used to control the number of channels in the block's output. Gaussian error linear units \cite[GELU;][]{gelu} are used as activation functions throughout the construction of the network, and layer normalisation \cite[LayerNorm;][]{layernorm} is used for normalisation.

\begin{figure*}
    \centering
    \includegraphics[width=0.7\textwidth]{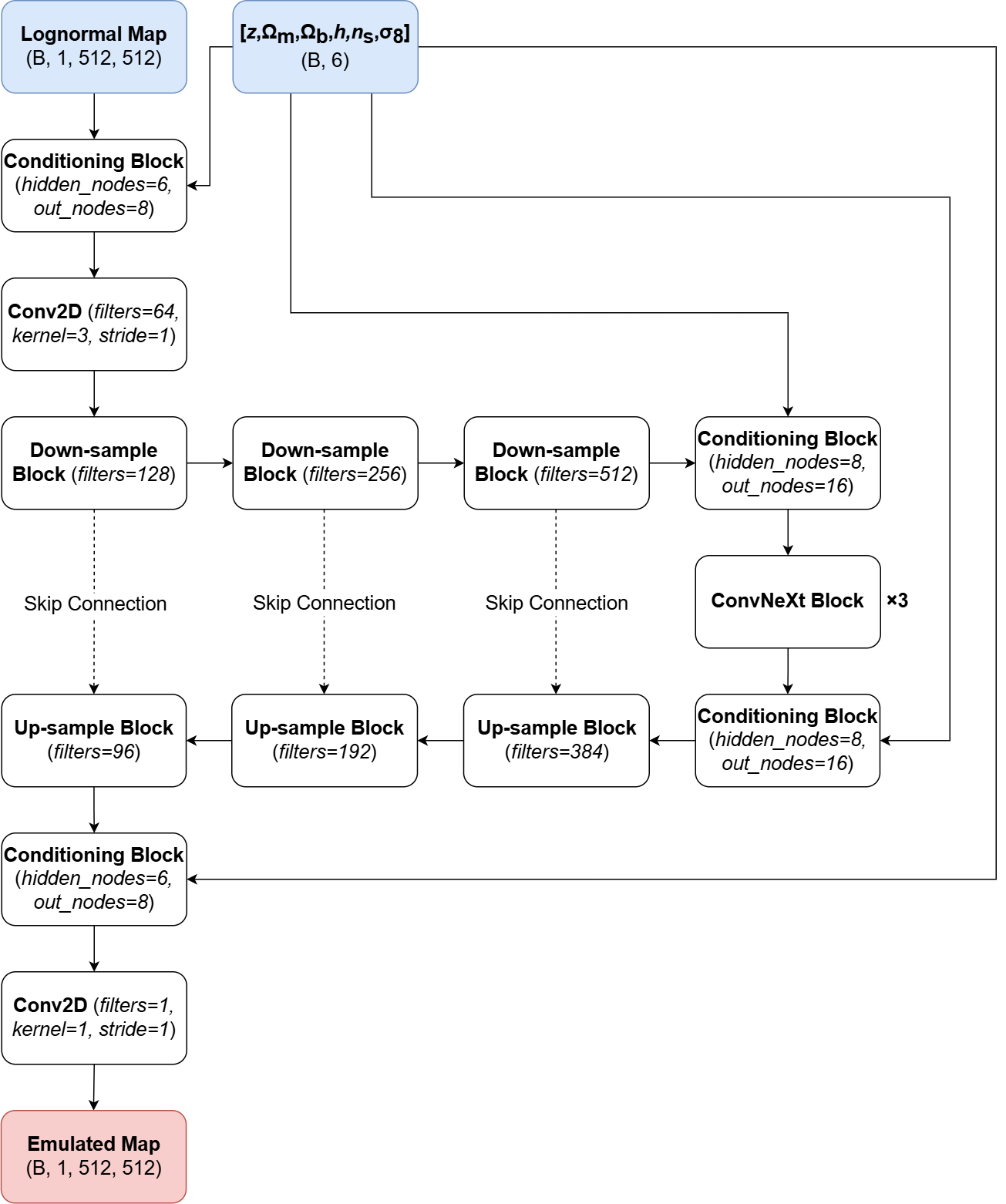}
    \caption{A schematic showing the construction of our conditional U-Net generator. We also provide information, in parentheses, regarding the dimensions for both the inputs (blue) and outputs (red) in roman font, and the hyperparameters of each layer in italics. Dimensions are quoted in the ``batch, channels, $\ast$'' convention, where $\ast$ represents any number of latent dimensions and B is used as a placeholder for the batch dimension of all inputs and outputs. All convolutional layers use circular padding in order to maintain the height and width of the input.}
    \label{fig:generator}
\end{figure*}

The construction of the generator can be found in Figure~\ref{fig:generator}. The generator consists of an initial $3 \times 3$ convolution followed by three down-sample blocks. We then introduce a bottleneck consisting of three ConvNeXt blocks, before using three up-sample blocks to return the input to its original resolution. We use a final $1 \times 1$ convolution to reduce the number channels back to $1$. We use conditioning blocks to inject information about redshift and cosmology before the initial and final convolutions, as well as before and after the bottleneck.

\bsp	
\label{lastpage}
\end{document}